\documentclass[oldversion,paper]{aa} 

\usepackage{natbib}

  \usepackage{subfigure}

\usepackage{graphicx}

\usepackage{txfonts}

\begin{document}

\title{An algorithm for correcting CoRoT raw light curves}

   \authorrunning{D. Mislis et. al.}

   \author{D. Mislis            \inst{1}$^,$\inst{2},
          J.H.M.M. Schmitt            \inst{1},  L. Carone \inst{3}, E. W. Guenther \inst{4} \& M. P\"{a}tzold \inst{3} }

\institute{$^{1}$Hamburger Sternwarte, Gojenbergsweg 112, D-21029 Hamburg, Germany \\
$^{2}$Institute of Astronomy, University of Cambridge, Madingley Road, Cambridge, CB3 0HA, United Kingdom \\
    $^{3}$ Rheinisches Institut f\"ur Umweltforschung, Abteilung Planetenforschung, an der Universit\"at K\"oln, Aachener Str. 209, 50931 K\"oln, Germany\\
    $^{4}$Th\"{u}ringer Landessternwarte Tautenburg, Sternwarte 5, D-07778 Tautenburg, Germany   \\
              \email{mdimitri@hs.uni-hamburg.de}
             }

\date{Received / Accepted }


\abstract{
We introduce the CoRoT detrend algorithm (\textit{CDA}) for detrending CoRoT stellar light curves. 
The algorithm \textit{CDA} has the capability to remove random jumps and systematic trends 
encountered in typical CoRoT data in a fully automatic fashion. Since enormous jumps in 
flux can destroy the information content of a light curve, such an algorithm is essential.
From a study of 1030 light curves in the CoRoT IRa01 field, we developed
three simple assumptions which upon CDA is based. We describe the algorithm analytically and provide some examples of how it works.
We demonstrate the functionality of the
algorithm in the cases of CoRoT0102702789, CoRoT0102874481, CoRoT0102741994, and CoRoT0102729260. Using CDA
in the specific case of CoRoT0102729260, we detect a candidate exoplanet around the host star of
spectral type G5, which remains undetected in the raw light curve, and estimate the planetary parameters
to be $R_{p}=6.27R_{E}$ and $P=1.6986$ days.}

   \maketitle
%


\section{Introduction}

The CoRoT satellite was successfully launched in 2006. Onboard CoRoT, there is a small 27cm telescope feeding two science channels to study astroseismology and transits, respectively 
\citep{2000JApA...21..319B}. The CoRoT has a field of view (FOV) of $\sim$ $2.7^{o}$ $x$ $3.05^{o}$. In its first observed field 
(IRa01 - $\alpha$ = $6^{h}46^{m}53^{s}$ \& $\delta$ = $-00^{o}12^{'}00{"}$), CoRoT observed continuously for 60 days, producing uninterrupted light curves for the first time. The data for 
the IRa01 field have been public since December 2008 and the astronomical community has access to these data.  Unfortunately, the  CoRoT light curves are affected by a variety of instrumental problems 
that severely hamper the data interpretation. To overcome these difficulties, we have developed the CoRot detrend algorithm (\textit{CDA}). In this paper, the algorithm is presented and 
we demonstrate its functionality on some typical CoRoT data sets.

\section{CoRoT light curves: The problems}



    \begin{figure}

    \centering

    \includegraphics[width=7.0cm]{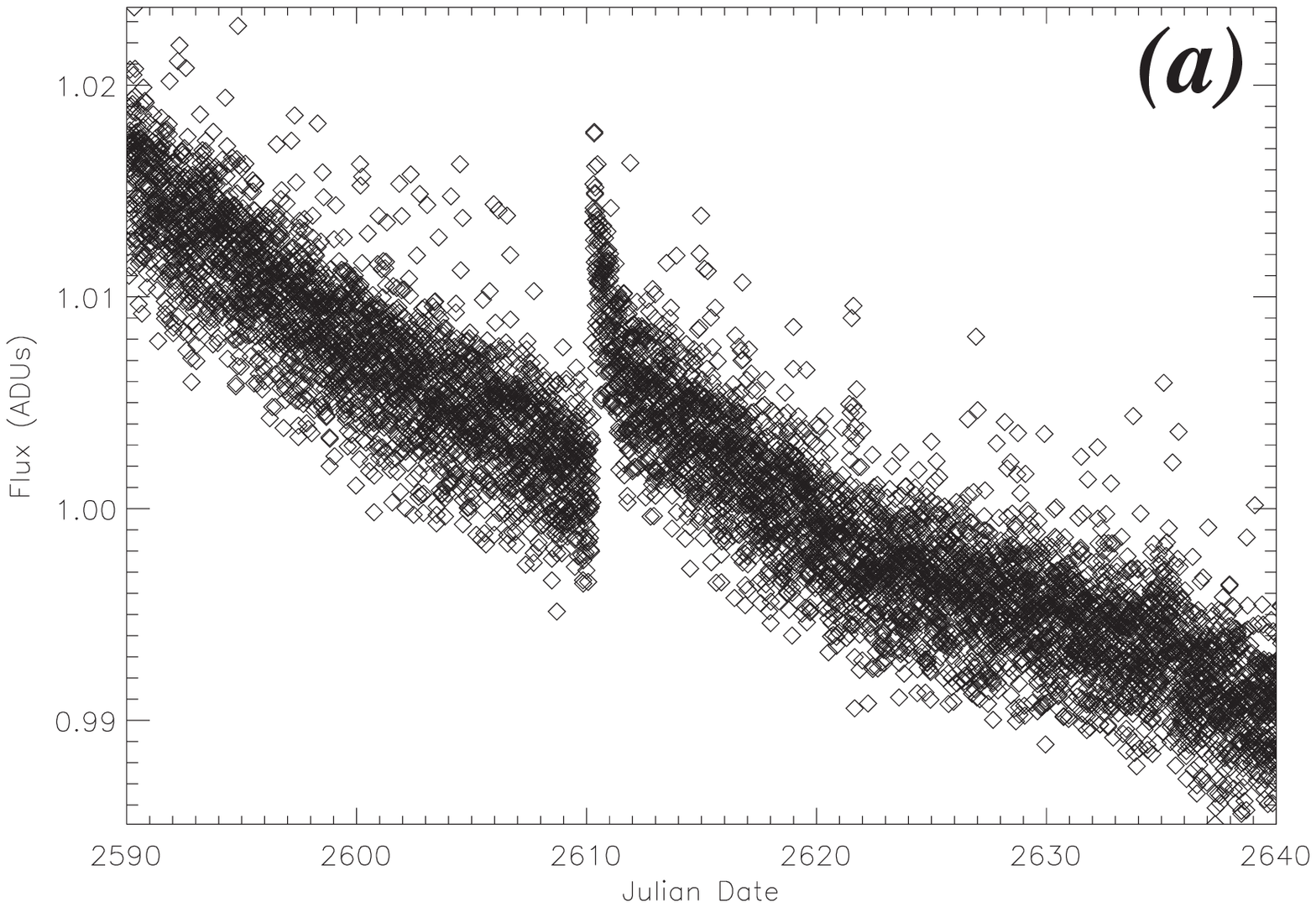}\\
    \includegraphics[width=7.0cm]{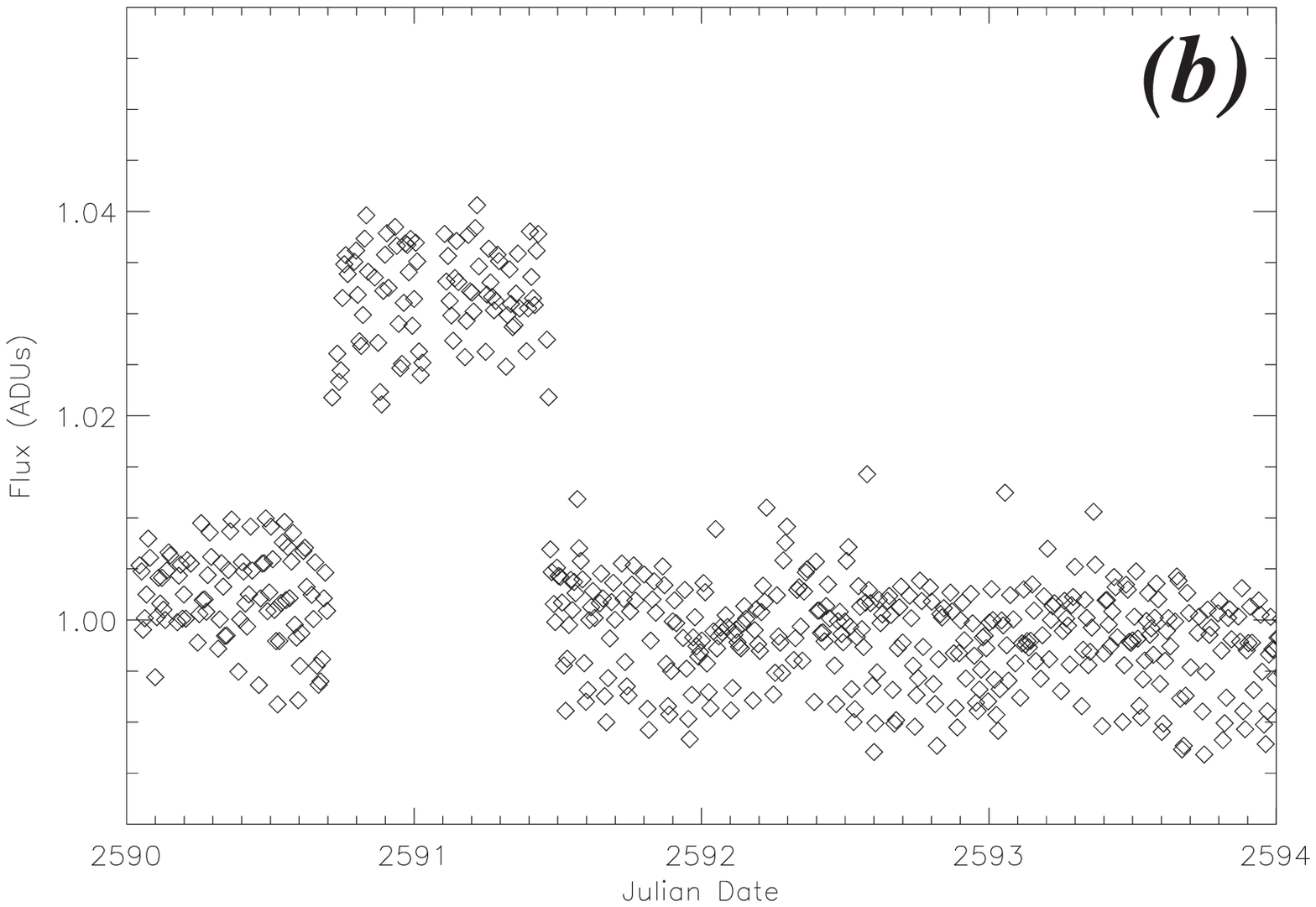}\\
   \includegraphics[width=7.0cm]{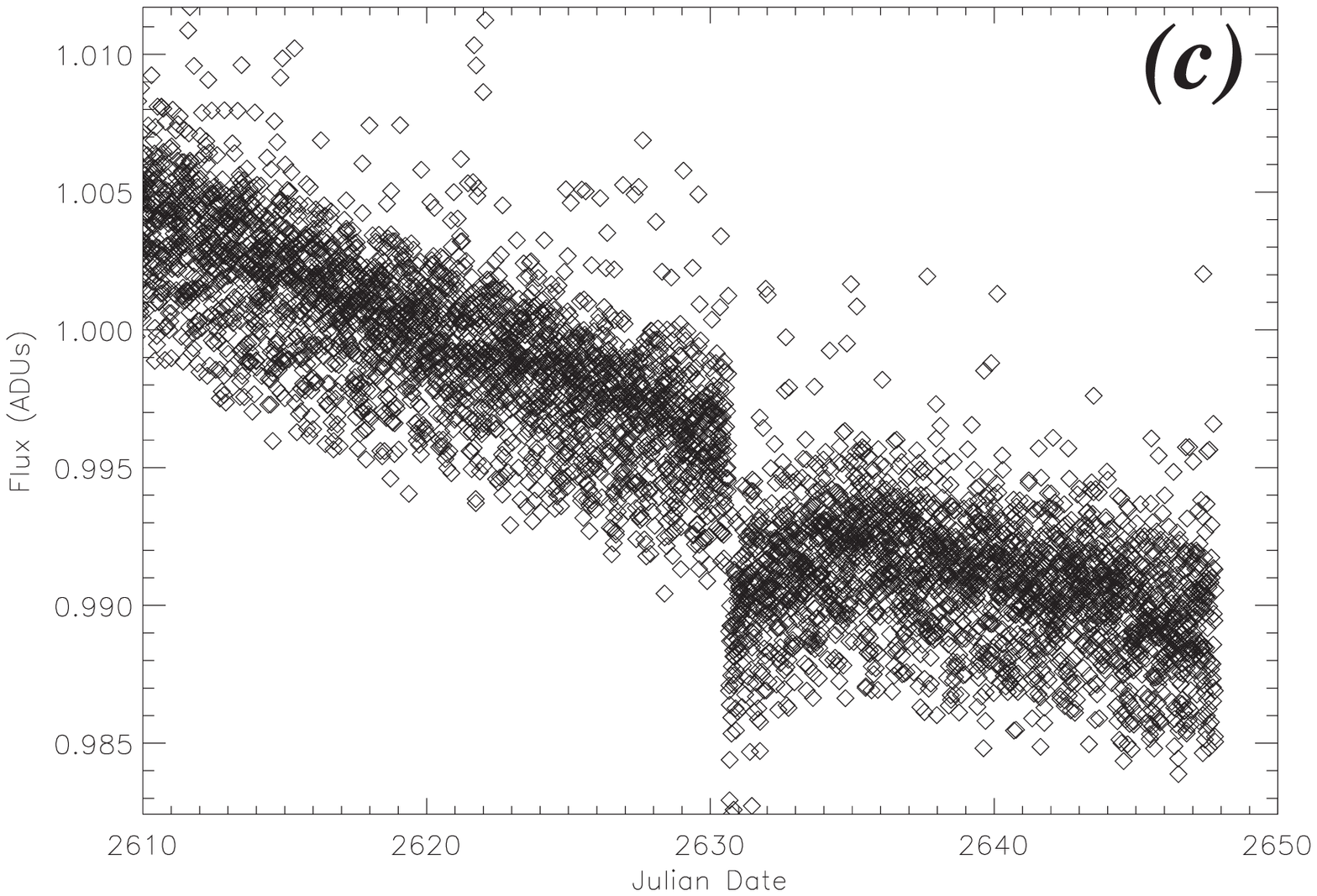}\\
    \includegraphics[width=7.0cm]{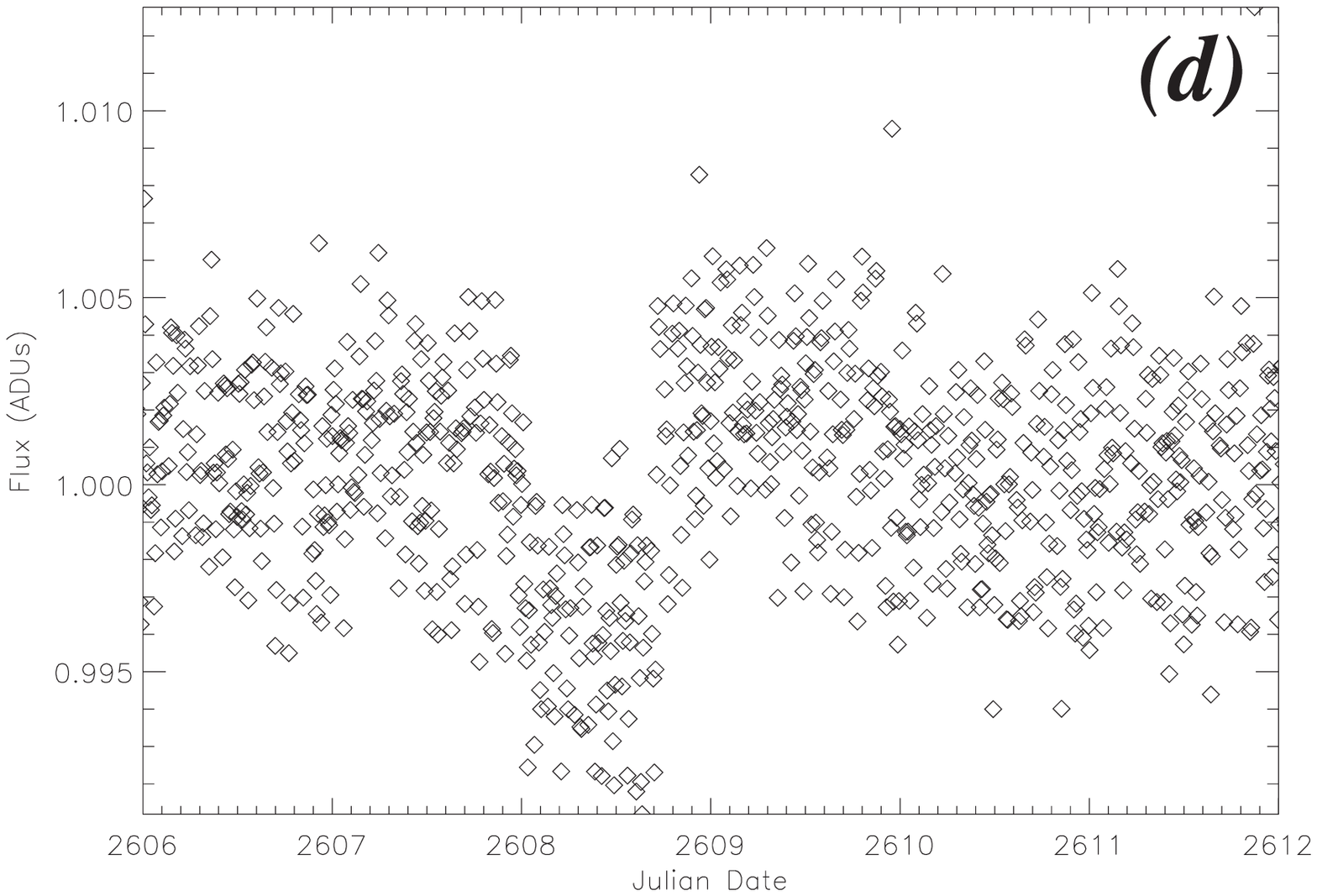}\\
   \includegraphics[width=7.0cm]{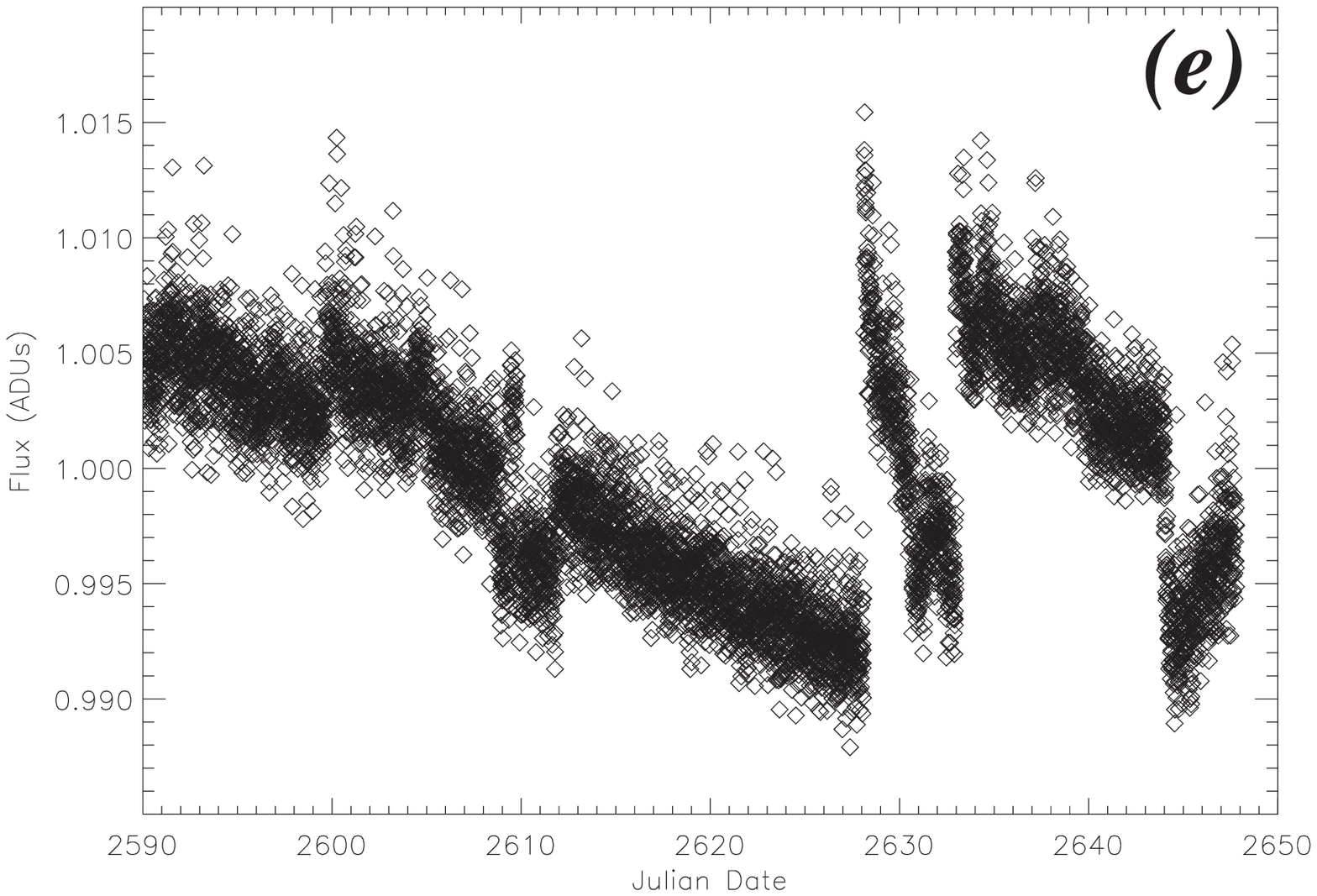}

   \caption{Jumps and trends in CoRoT light curves. CoRoT01027-21492 \textbf{(a)} -24482 \textbf{(b)} -40879 \textbf{(c)} -49307 \textbf{(d)} - 27431 \textbf{(e)}. }
\label{fig1}
\end{figure}

The CoRoT data files contain multi-color light curves that were produced by inserting a low-resolution dispersing prism into 
the telescope beam. This set-up is desinged to provide  simultaneous light curves in the red (R), 
green (G), and blue (B) bands, although, these bands do not correspond to true photometric filters and, the bands may indeed 
differ from star to star. In this paper, we study the multi-color data, but also consider the total (white) flux obtained
by summing up the individual light curves through $W=R+G+B$.

Figure \ref{fig1} are typical CoRoT light curves from IRa01. The first panel of Fig. \ref{fig1} shows a typical exponential jump very similar to a flare star. A trend is also evident. In the second light 
curve, there appears to be a box-shape jump, and in the third and fourth light curves one can discern features similar to those in the first and second light curves, except that the jumps are downwards.
We note that the downward jump  in the third light curve is very similar to a transit event, thus making the detection of true transits difficult. Combinations of all the above features appear to 
produce a rather typical CoRoT light curve. There are, two basic instrumental problems with all CoRoT light curves. First, 
there is a long-term trend, forcing a secular decrease in the light curve intensity over the full observing period of 60 days. The strengths of the trends in different sources may be 
different; the physical cause of these trends is not well understood. The second and even more serious problem 
is the instrumental jumps in the light curves. The term ``jump'' refers to a sudden variation in intensity without any obvious reason. Many of these jumps resemble stellar flares, although, the vast 
majority are clearly instrumental nature. Their physical explanation could be cosmic radiation and the time evolution of bright pixels \citep{2008MNRAS.384.1337P}. These jumps are a random phenomenon 
and affect each filter differently. An inspection of hundreds of CoRoT light curves similar to those presented in Fig. \ref{fig1} allows us to classify the observed shapes of jumps into five groups:

\begin{itemize}
\item Sudden intensity increase and exponential decrease (Fig. \ref{fig1} - panel a); 
\item Sudden intensity increase and decreases (box shape, Fig. \ref{fig1} - panel b);
\item Sudden intensity decrease and exponential increase  afterwards (Fig. \ref{fig1} - panel c);
\item Sudden intensity decrease and increase (negative box shape, Fig. \ref{fig1} - panel d);
\item All of the combinations above (Fig. \ref{fig1} - panel e).
\end{itemize}

A statistical analysis of IRa01 field (visual inspection) shows that only a small minority (Table \ref{tab1}) of all jumps are powerful enough to simultaneously appear in each colour.
Most of the light curves are affected not only by one single jump, but by many jumps occurring in the different filters at different times. In Table \ref{tab1}, we show the results of a statistical study 
of the appearance and the shapes of jumps using data for IRa01. The three first columns of Table \ref{tab1} show the number of light curves affected by jumps in the respective filter filter and 
the fourth column indicates the total size of the effect.

  \begin{table}

\begin{center}

      \caption[]{Statistical analysis of 1030 CoRoT light curves from IRa01. Jumps appear in more than 50\% of all light curves in all filters; in 0.82\% of all 
light curves, jumps in all filters occur at the same time.}

         \label{tab1}

        $\begin{array}{llll}

            \hline

            \noalign{\smallskip}

            R$ $filter & G$ $filter & B$ $filter & Total \\

            \noalign{\smallskip}

             \hline

            \noalign{\smallskip}

            38.14 \% & 14.4 \% &  15.1 \% &  67.6 \%  \\

            \hline\hline

         \end{array}$

\end{center}

   \end{table}

\section{The CDA Algorithm}

\subsection{General features}

\begin{figure}

    \centering

    \includegraphics[width=8.0cm]{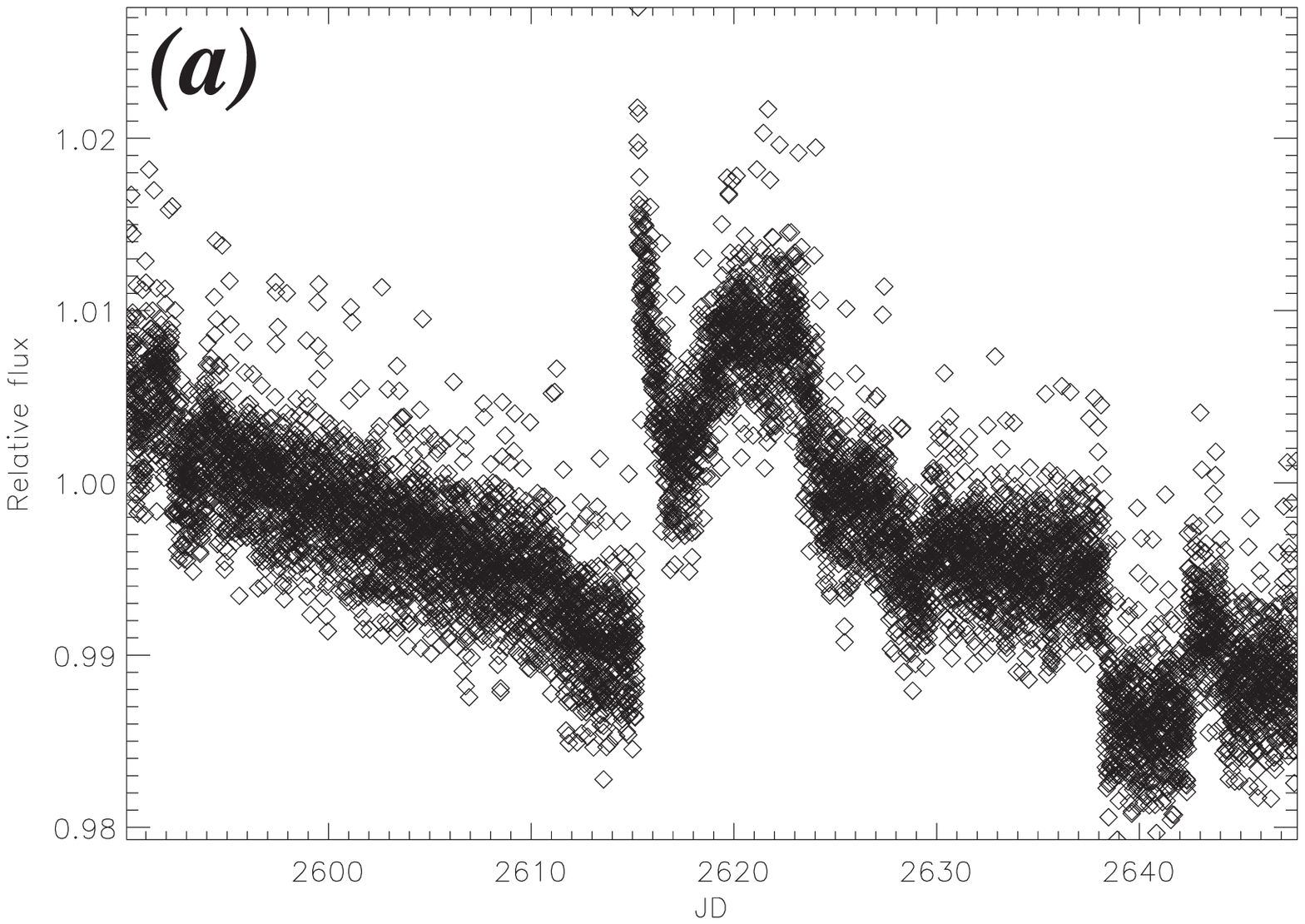}\\
    \includegraphics[width=8.0cm]{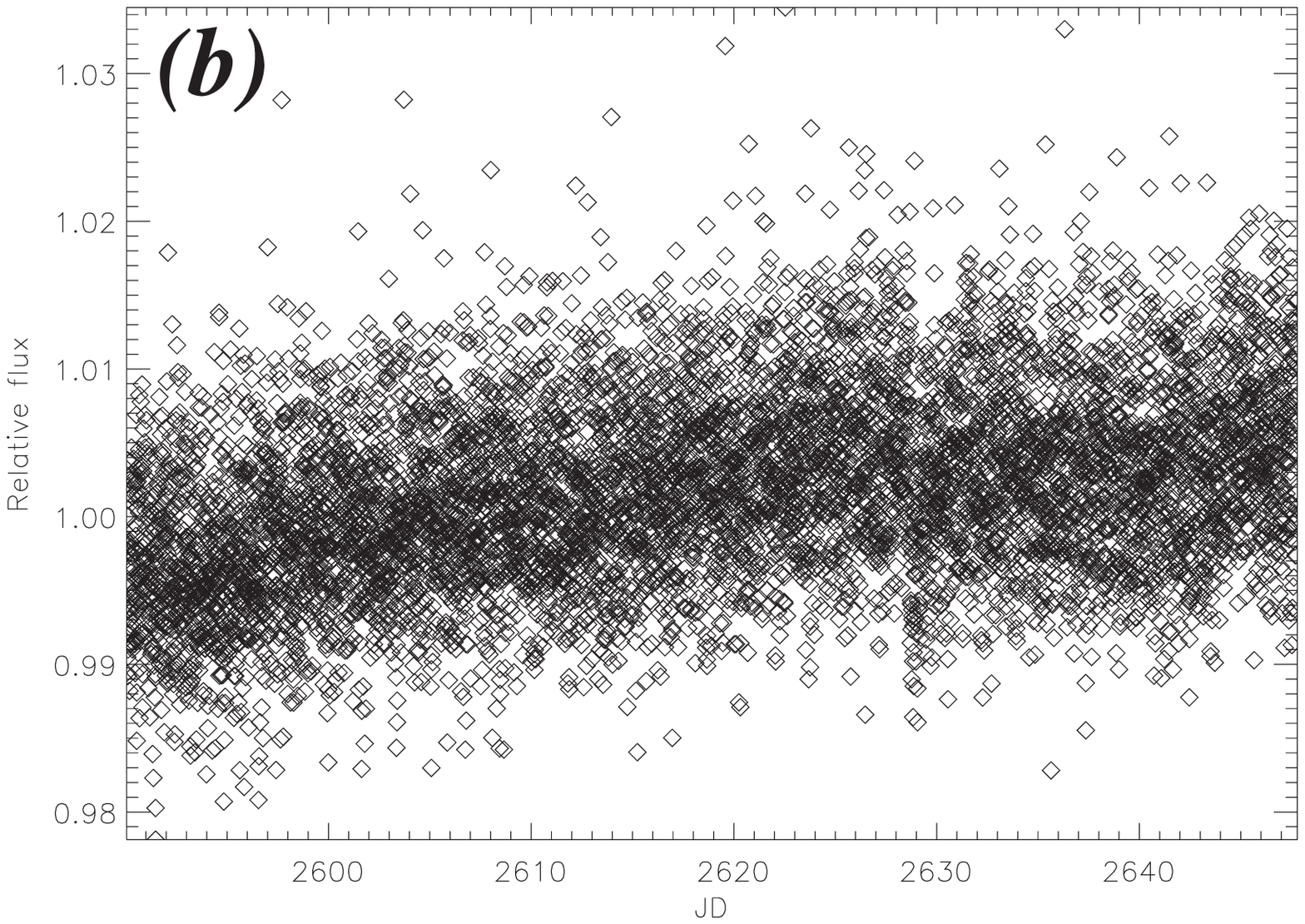}\\
   \includegraphics[width=8.0cm]{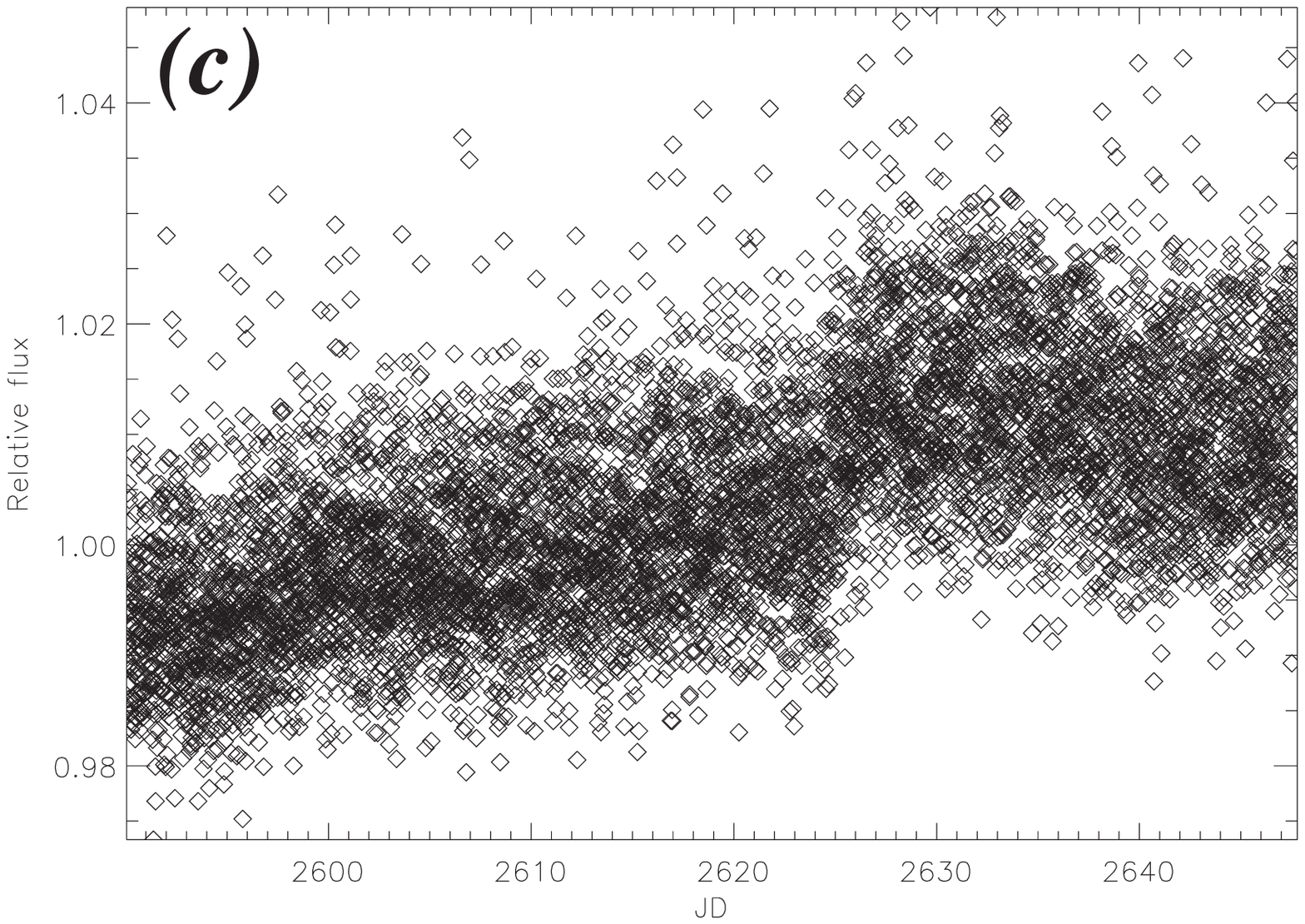}

   \caption{CoRoT0102729260. Three filter light curves (R \textbf{(a)}, G \textbf{(b)}, B \textbf{(c)}) from a data set. The jumps in the red light curve do not appear in the other 
filters and vice versa.}
\label{fig2}
\end{figure}

It is quite difficult to describe all the features perturbing a CoRoT light curve with a given function, since  there are many different shapes of jumps with many different functional forms.  Furthermore,
the problem is complex because we do not know which light curve features are real signals (real transits, real flares etc.) and which are instrumental effects. The algorithm is based on 
three assumptions. We first assume that trends appear in almost all light curves and both flux increases and decreases can occur. The trends are not periodic and we assume them to be a long-term phenomenon 
\citep{2009A&A...506..425A}. Our second assumption is based on the statistical analysis results for the data. The study of 1030 
light curves from IRa01 field shows that only 0.82 $\%$ of them are affected by a jump in all three filters at the same time. In these cases, the jump is very large and affects all bands with the same 
temporal pattern, although, in most cases the jumps affect only one band at any given time (Fig. \ref{fig2}).
We therefore ignore the cases in which jumps occur simultaneously in all three bands. Our third assumption is that real transits must appear in all three filters, while, of course, the intensity and 
transit depth can vary from filter to filter.  In summary, for the \textit{CDA} we assume that \par

\begin{itemize}
 \item Long-term trends appear in all CoRoT light curves;
 \item Jumps are random phenomena appearing in different filters at different times;
 \item The real signals from transits appear in all three bands.
\end{itemize}

We emphasize that \textit{CDA} works only for events (such transits) that appear in two or more bands. Hence \textit{CDA} does not work for stellar flares, since most stellar flares do not show any 
flux enhancements in the red and green band, but only in the blue band. In these circumstances, \textit{CDA} will destroy real signals, unless the flare is powerful enough to appear in all bands. 

\subsection{The algorithm}

To remove instrumental feature, \textit{CDA} uses simultaneously all the colour light curves. The basic idea of \textit{CDA} is to use the cleanest
filter band as a proxy for the whole light curve. The raw data files of each CoRoT light curve have a quality flag (CoRoT files - Col. 4), indicating the quality of each data point 
\citep{2009A&A...506..431M}. We first remove all the ''bad points'' (points flagged by CoRoT to have high noise); we note 
that these ''bad points'' are the same for all the filters of each star. In this paper, we use light curves from which all ''bad points'' have been removed (as in Fig. \ref{fig1}). As noted in our first 
assumption, trends are a long-term phenomenon. A $3^{rd}$ degree polynomial is fit to the entire light curve to remove the trend in each filter per star. Because each CoRoT light curve typically 
has thousands of data points, the polynomial does not fit short-term variations and real short-term events such as transits. We thus write 

\begin{equation}
Flux = a + b\cdot JD+c\cdot JD^{2}+d\cdot JD^{3}, 
\label{eq1}
\end{equation}
where JD is the Julian date (normalized to range $-1 \leq JD \leq 1$) and a, b, c, and d are the fit parameters for the third degree polynomial. At the end of this procedure, we have a detrended light 
curve per filter for each star. \par
\begin{figure}
\centering
\includegraphics[width=8cm,angle=0,clip=true]{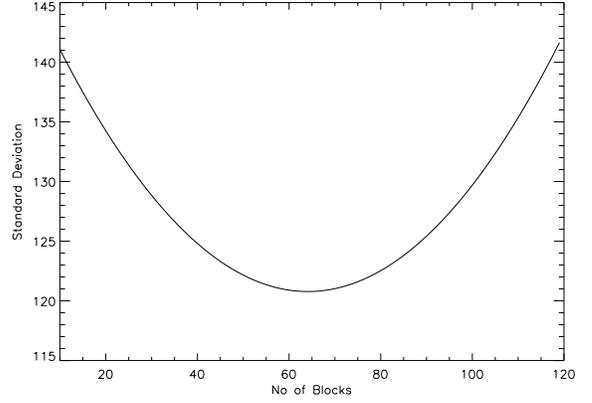}
\caption{Standard deviation versus number of blocks.}
\label{bloc}
\end{figure}

After this step, \textit{CDA} proceeds in removing the jumps. To identify the cleanest light curve for a reliable jump removal, we create ''sub-light curves'', which have a typical duration 
of a day. Thus, for the IRa01 field we create 60 ''sublight'' curves, called simply light curves in the following. These 60 blocks were selected after we checked various combinations. If the number 
of blocks is too small, the probability of including a jump in the ''sublight'' curve increases. Figure \ref{bloc} shows the optimal block number versus standard deviation. \\

We assume that there are three full light curves for a given star in each band with $N$ points per light curve which we denote by $F_{R,i}$, $F_{G,i}$, and $F_{B,i}$ with $i=1,N$ the individual data 
values in the red, green, and blue filters, respectively. We then divide each color light curve into 60 sub-light curves (one sub-light curve per day for IRa01 - 60 days). For each sub-light curve, we 
calculate the mean value $MR$, $MG$, and $MB$ and normalize each sub-light curve by its mean value. We then compute new, normalized sub-light curves $NF$ to be

\begin{equation}
NF_{R,G,B,i} = \frac{F_{R,G,B,i}}{M_{R,G,B}} ,
\label{eq2}
\end{equation}
for each filter band, and it is clear that all of these light curves have a mean of unity.  This normalization is necessary otherwise the entire process would be dominated by the light curve of the 
strongest signal, which is usually the red light curve. As a side effect, \textit{CDA} normalizes the depth of a possible transit in all filters using Eq. \ref{eq2}, so when the algorithm continues 
with its next steps, all transit events in each filter will have the same depth,thus \textit{CDA} does not destroy real signals from the transits. 

\begin{figure*}
\centering

\includegraphics[width=7cm,angle=0,clip=true]{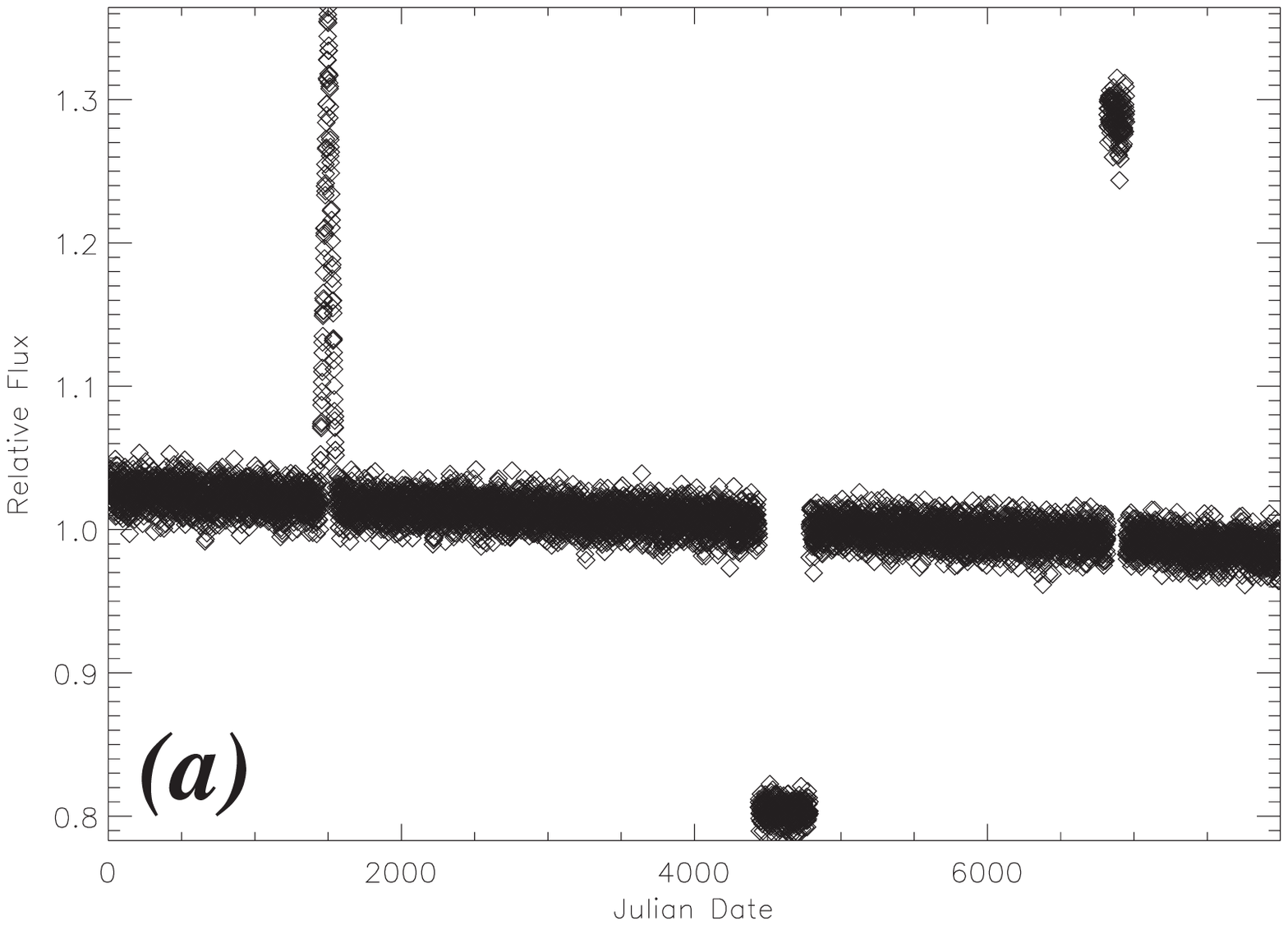}
\includegraphics[width=7cm,angle=0,clip=true]{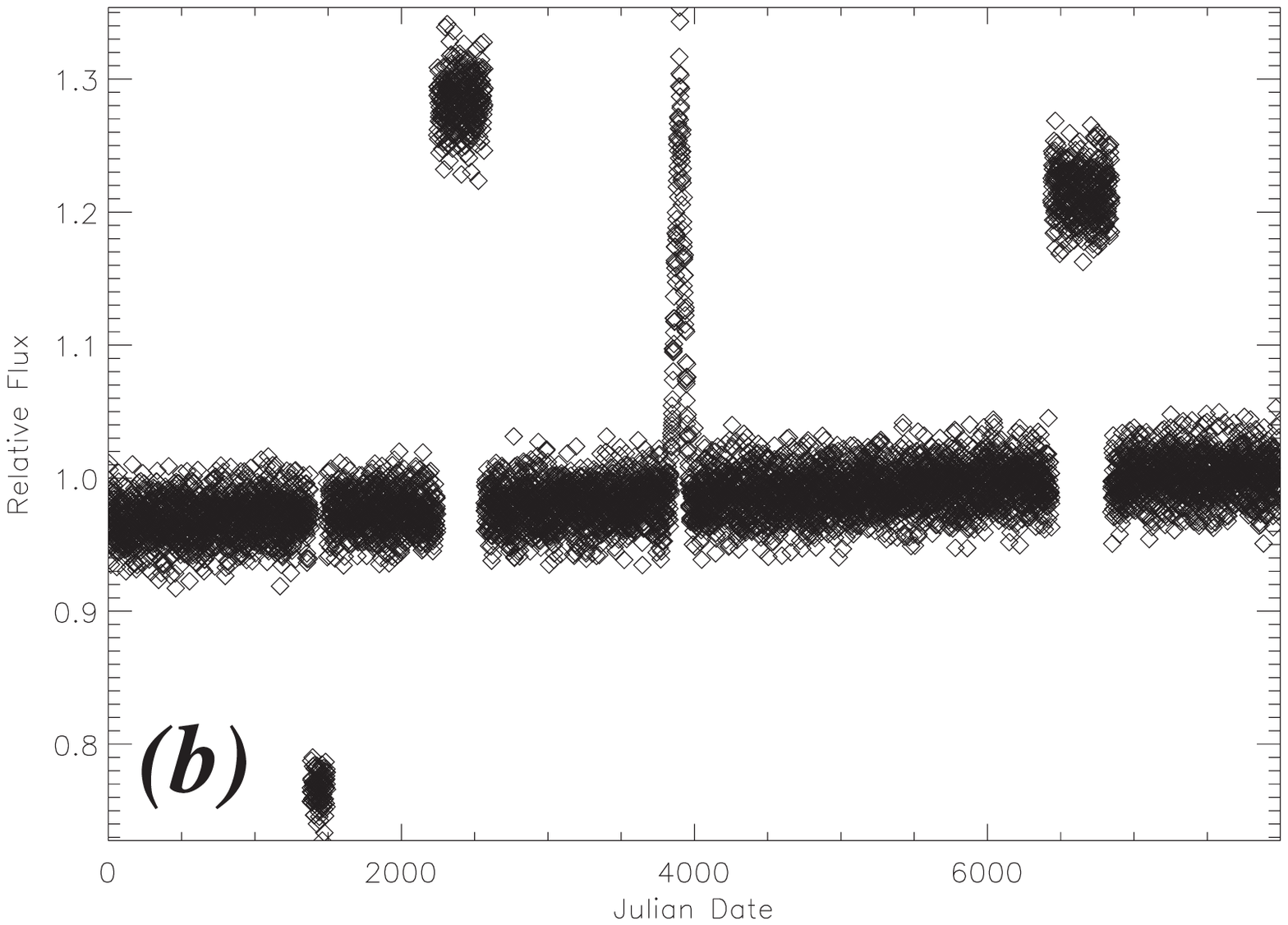}
\includegraphics[width=7cm,angle=0,clip=true]{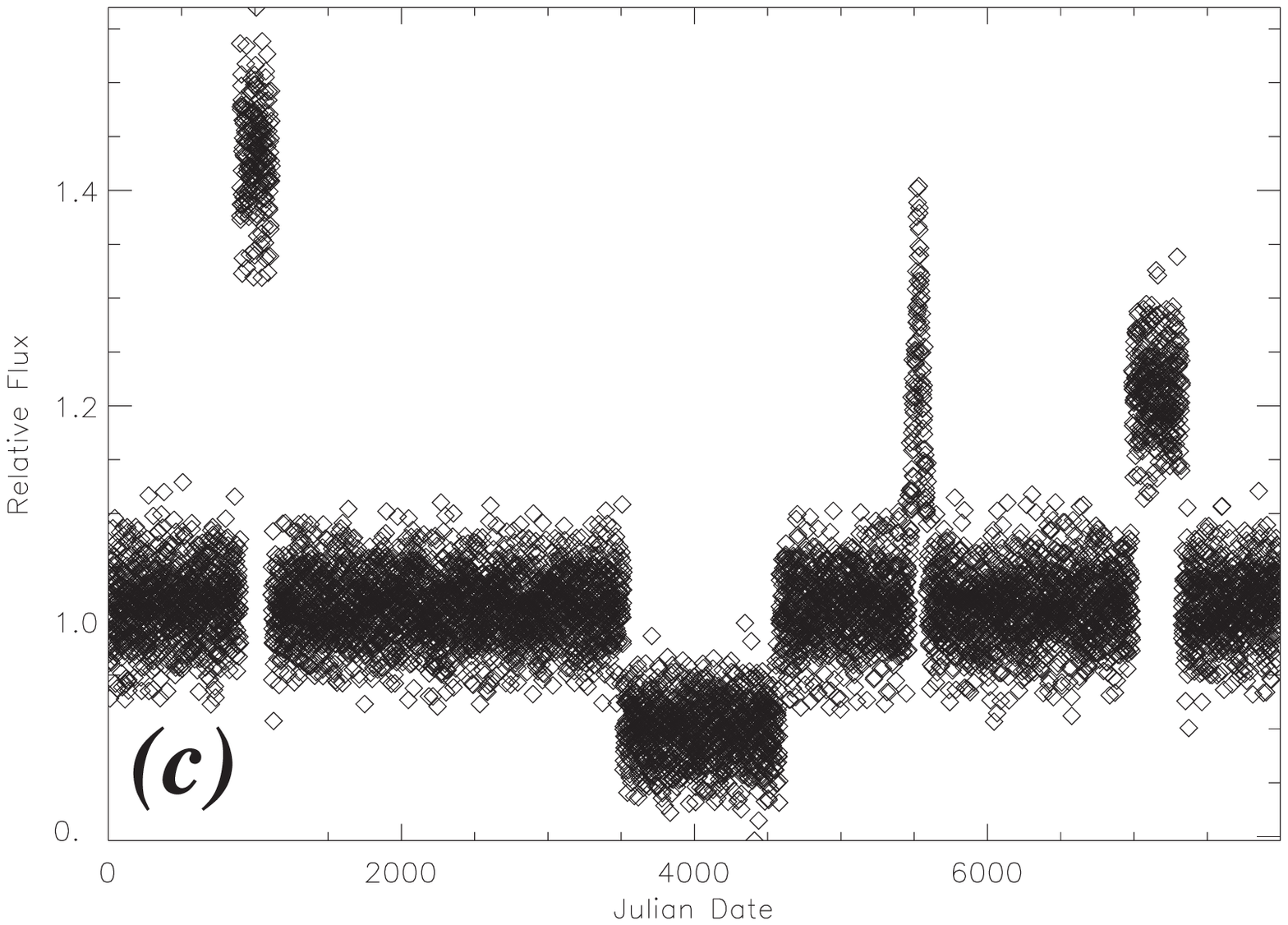} \\
\includegraphics[width=7cm,angle=0,clip=true]{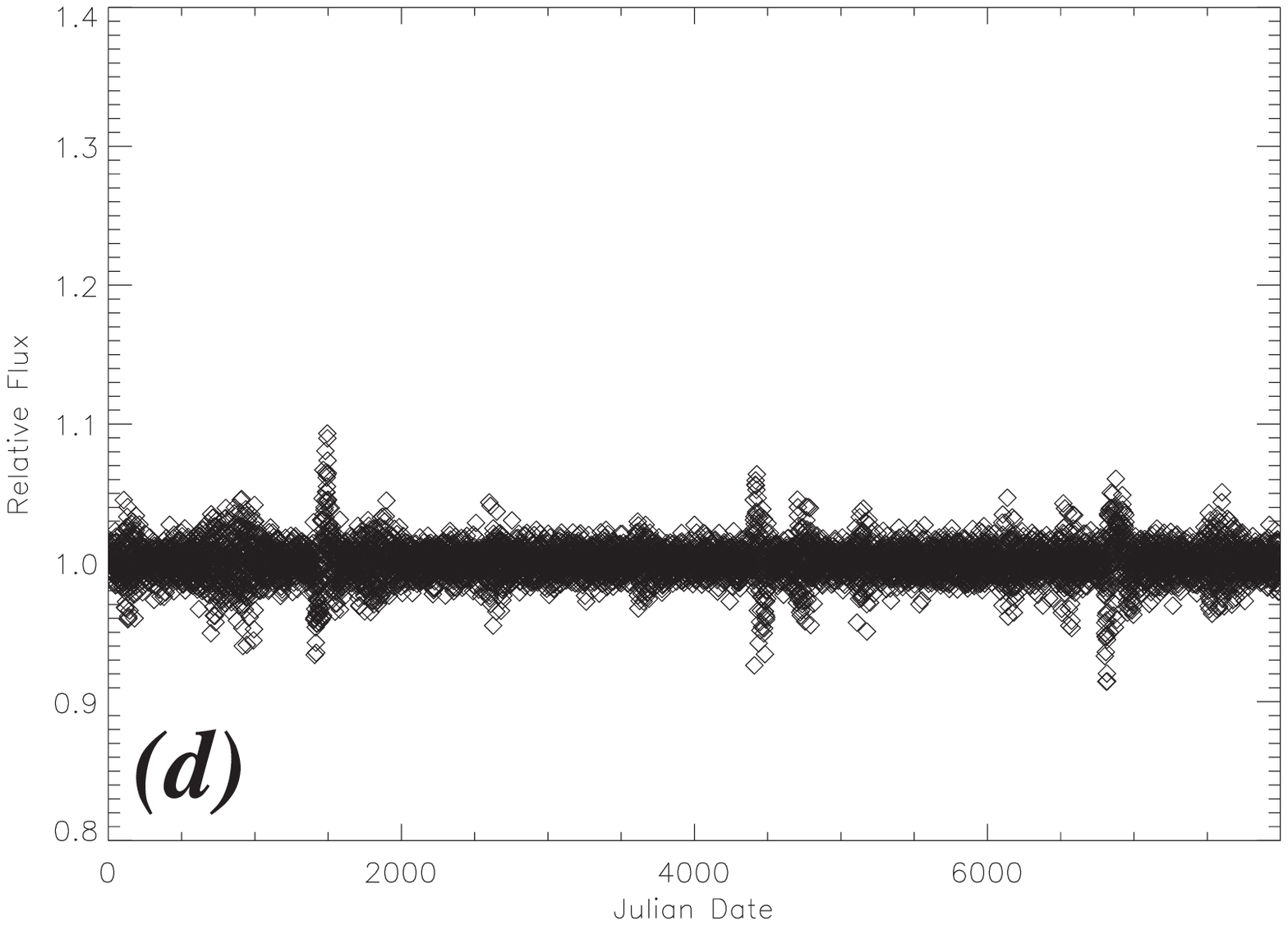}
\includegraphics[width=7cm,angle=0,clip=true]{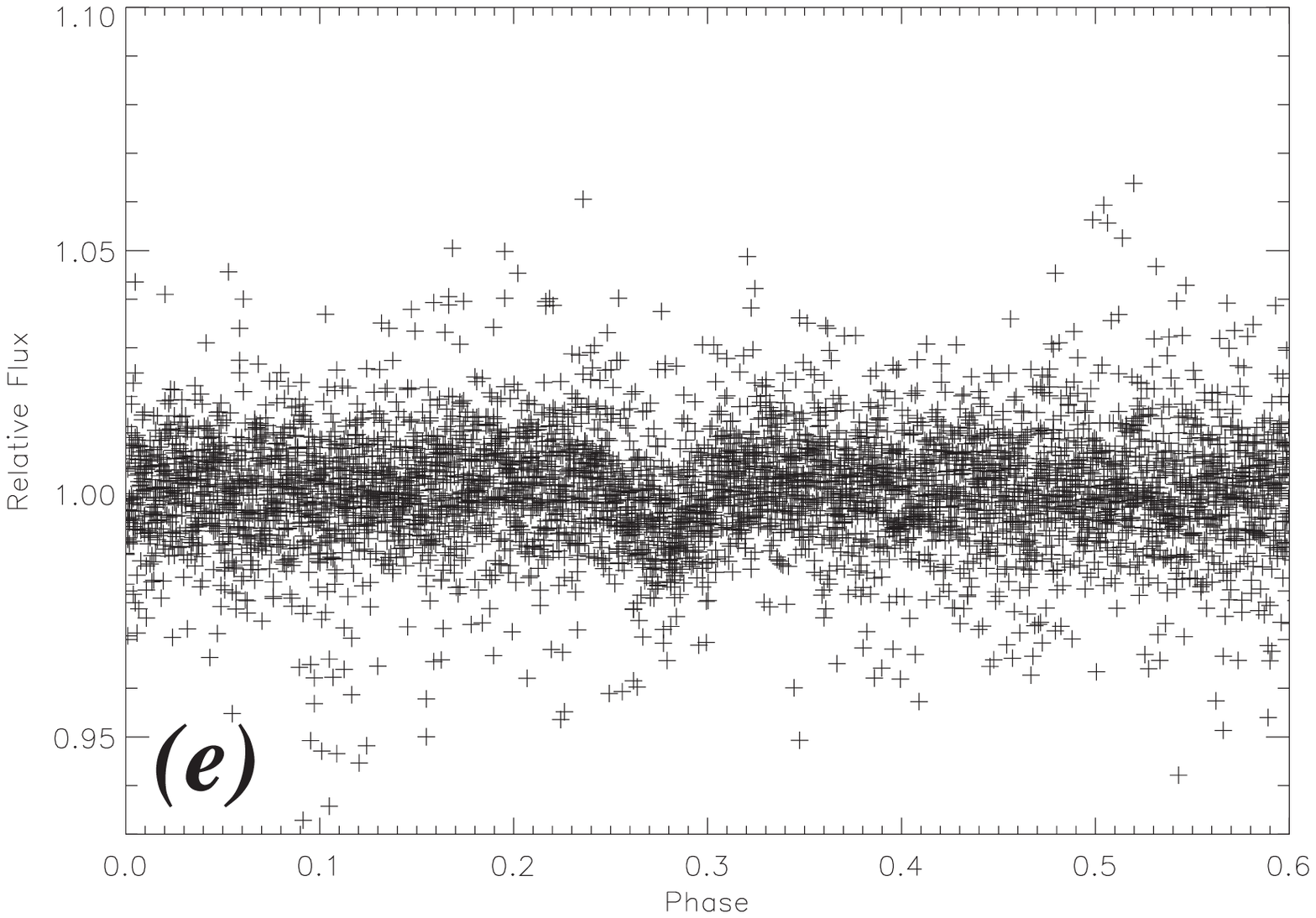}

\caption{Simulated data. R - \textbf{(a)}, G - \textbf{(b)} and B -\textbf{(c)} color, respectively. Plot \textbf{(d)} is the final light curve after CDA and the plot \textbf{(e)} is the phase diagram of 
the transit after CDA \& BLS.}
\label{sim1}
\end{figure*}

The normalized light curves now have the same mean, but their dispersions, differ. Our next goal is to identify the instrumental scatter caused, for example 
by jumps in each light curve and differentiate this instrumental scatter from statistical noise. To achieve this, \textit{CDA} extracts five random packages of twenty adjacent points each from 
all colour bands and calculates the standard deviation of each package per filter; the result should represent a good estimate of the correct light curve value at that time. If we use many packages, the 
probability of including jumps increases. The correct combination of packages and points is a function of the duration of the jumps, which is a random value, thus there is no ideal combination. We 
define as the mean standard deviation $(MSD)$ the mean value of these five packages for each filter

\begin{equation}
MSD_{R,G,B}=\frac{1}{5}\displaystyle\sum_{j=1}^{5} \frac{1}{20}\sqrt{\displaystyle\sum_{i=k_j}^{k_j+20} (NF_{R,G,B,i}-Mean_{min})^2},  
\label{eq4}
\end{equation}
where $k_j$ denotes 5 different random data points of the light curve and $Mean_{min}$ is the mean value of the flux of each package. In general, each filter has a different $MSD$ 
value, which is compared with the standard deviation of each filter $TSD$ defined to be

\begin{equation}
TSD_{R,G,B} = \frac{1}{N}\sqrt{\displaystyle\sum_{i=1}^N (NF_{R,G,B,i}-Mean_{min})^2}. 
\label{eq5}
\end{equation}

Finally, the relative standard deviation of each filter \textit{RSD} is computed and defined to be 

\begin{equation}
RSD_{R,G,B} = \frac{TSD_{R,G,B}}{MSD_{R,G,B}}.
\label{eq6}
\end{equation}
At the end of this process, we have three normalized light curves $NF_{R,i}$, $NF_{G,i}$, and $NF_{B,i}$, and three values for the relative standard deviation $RSD_{R}$, $RSD_{G}$, and $RSD_{B}$ for each 
filter light curve, respectively. The algorithm \textit{CDA} compares these three numbers and refers to the light curve with the minimum \textit{RSD} as the base and the light curve with the maximum 
\textit{RSD} as the target. To help illustrate the procedure, we continue with an example. We assume that the base is the blue light curve $(NF_{B,i})$ and the target is the red $(NF_{R,i})$ light curve. 
Using both the base and the target, \textit{CDA} calculates a new mean light curve $(AF_{i})$; in our example, \textit{CDA} computes

\begin{equation}
AF_{i} = \frac{1}{2}(NF_{R,i}+NF_{B,i}).
\label{eq7}
\end{equation}
and then refers to $AF_{i}$ as the light curve with the maximum \textit{RSD} (in this example, it defines $AF_{i}$ to be $NF_{R,i}$). According to assumptions 2 and 3, in the $AF_{i}$ light curve 
any possible real signal remains but all the fake (jump) signals tend to be reduced, because jumps appear only at specific times in each filter. As a final result, we have in our example a red light 
curve detrended and two others (green and blue) untouched. If we try to run the algorithm again, we notice that the new values of \textit{RSD} have changed because one light curve has changed. This means that every time we run the 
previous step of the algorithm, \textit{CDA} removes a part of a fake signal (Fig. 3). \par
When these loops end, we renormalize the final light curve of the red channel to the raw mean value

\begin{equation}
NFR_{final} = NR \cdot NF_{R,i} ,
\label{eq8}
\end{equation}

\begin{figure*}
\centering

\includegraphics[width=8cm,angle=0,clip=true]{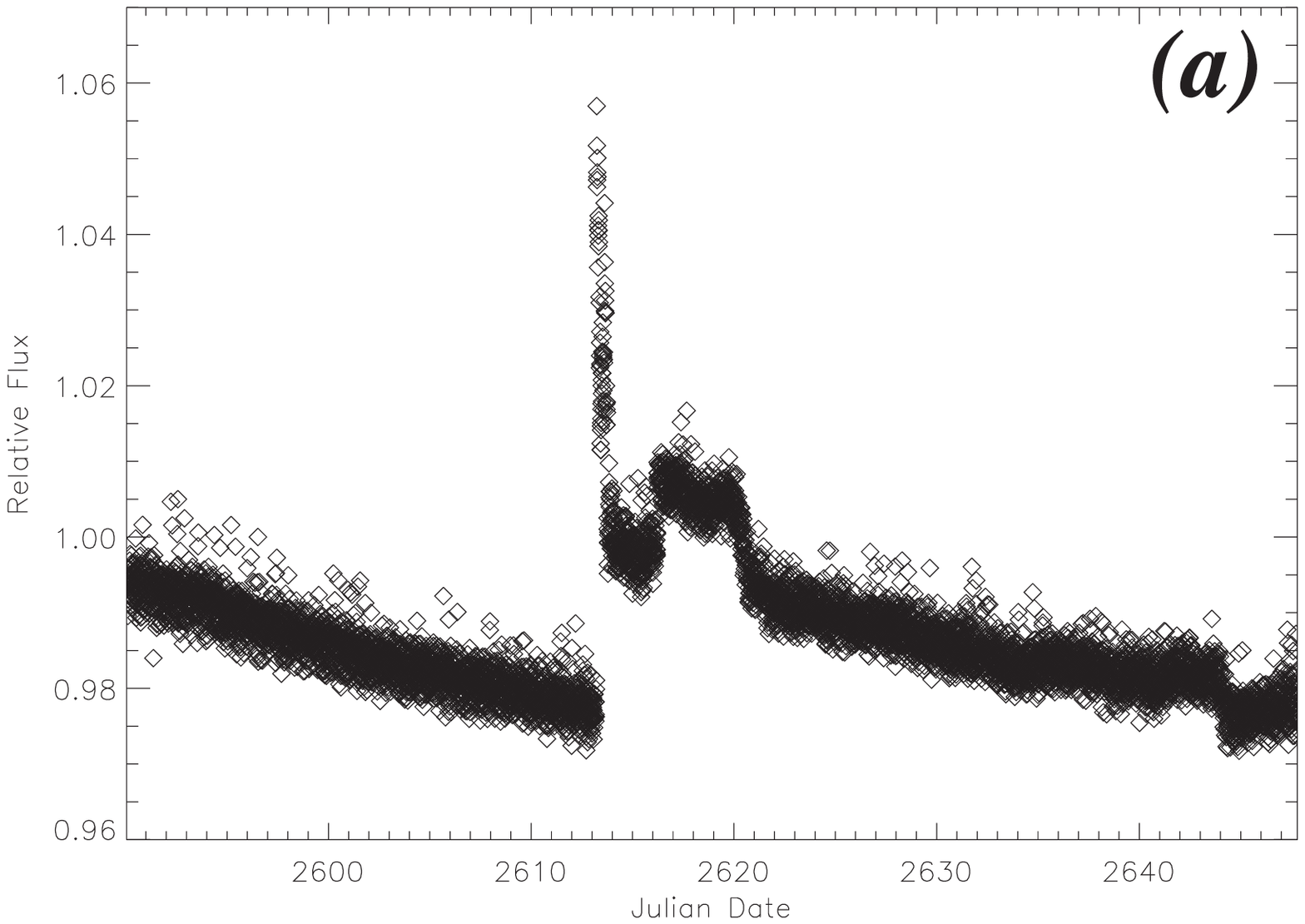}
\includegraphics[width=8cm,angle=0,clip=true]{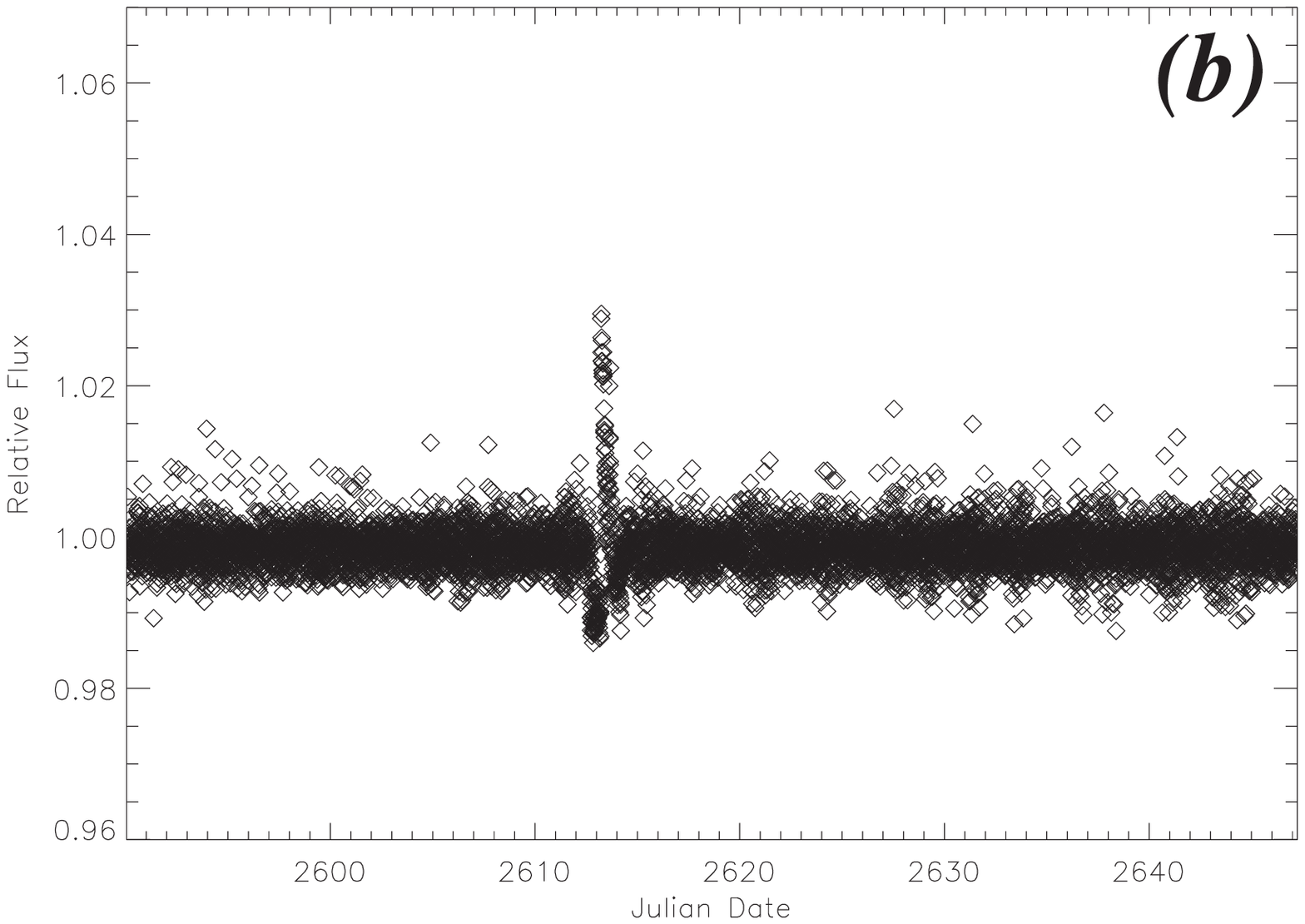} 
\includegraphics[width=8cm,angle=0,clip=true]{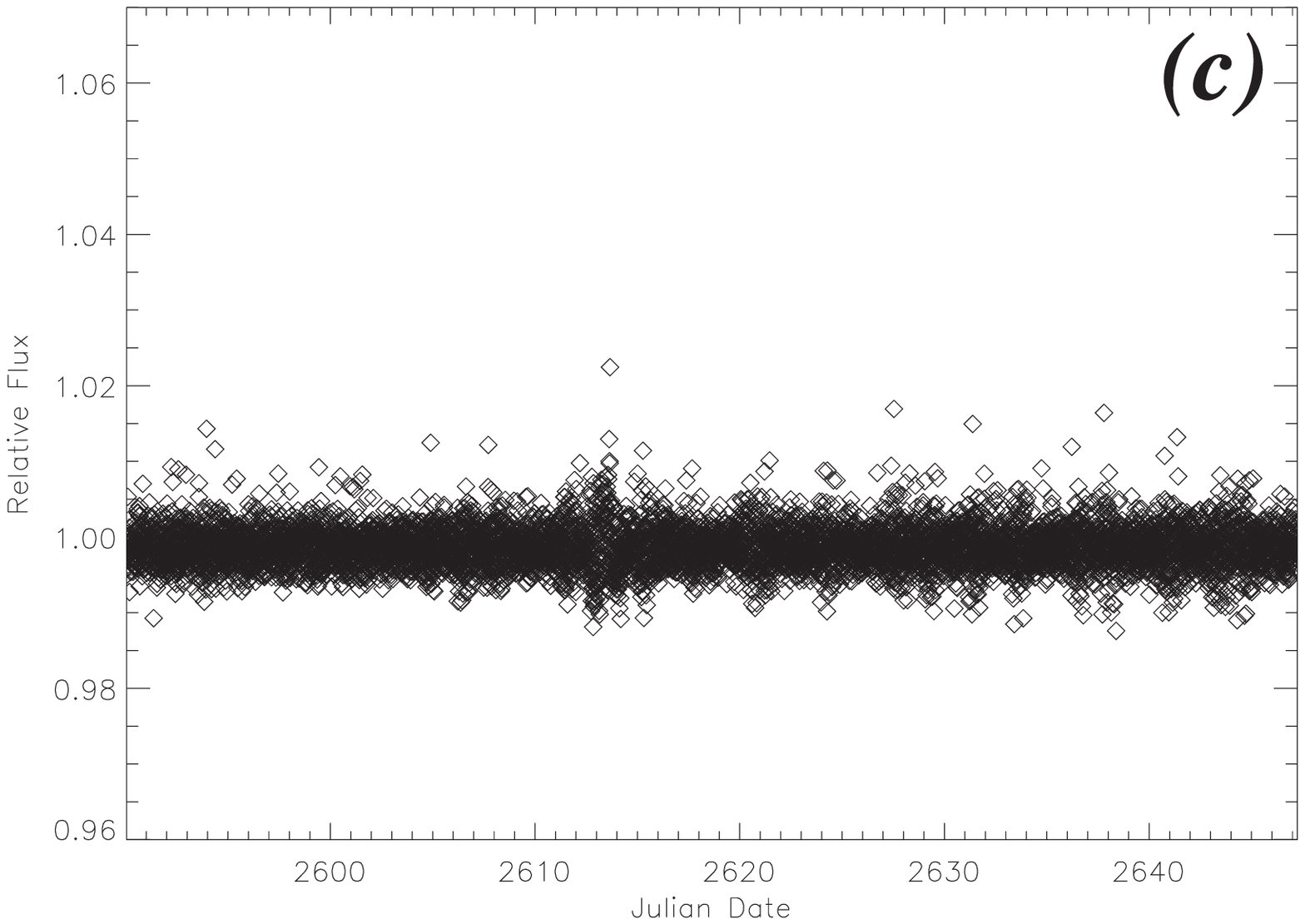} 
\includegraphics[width=8cm,angle=0,clip=true]{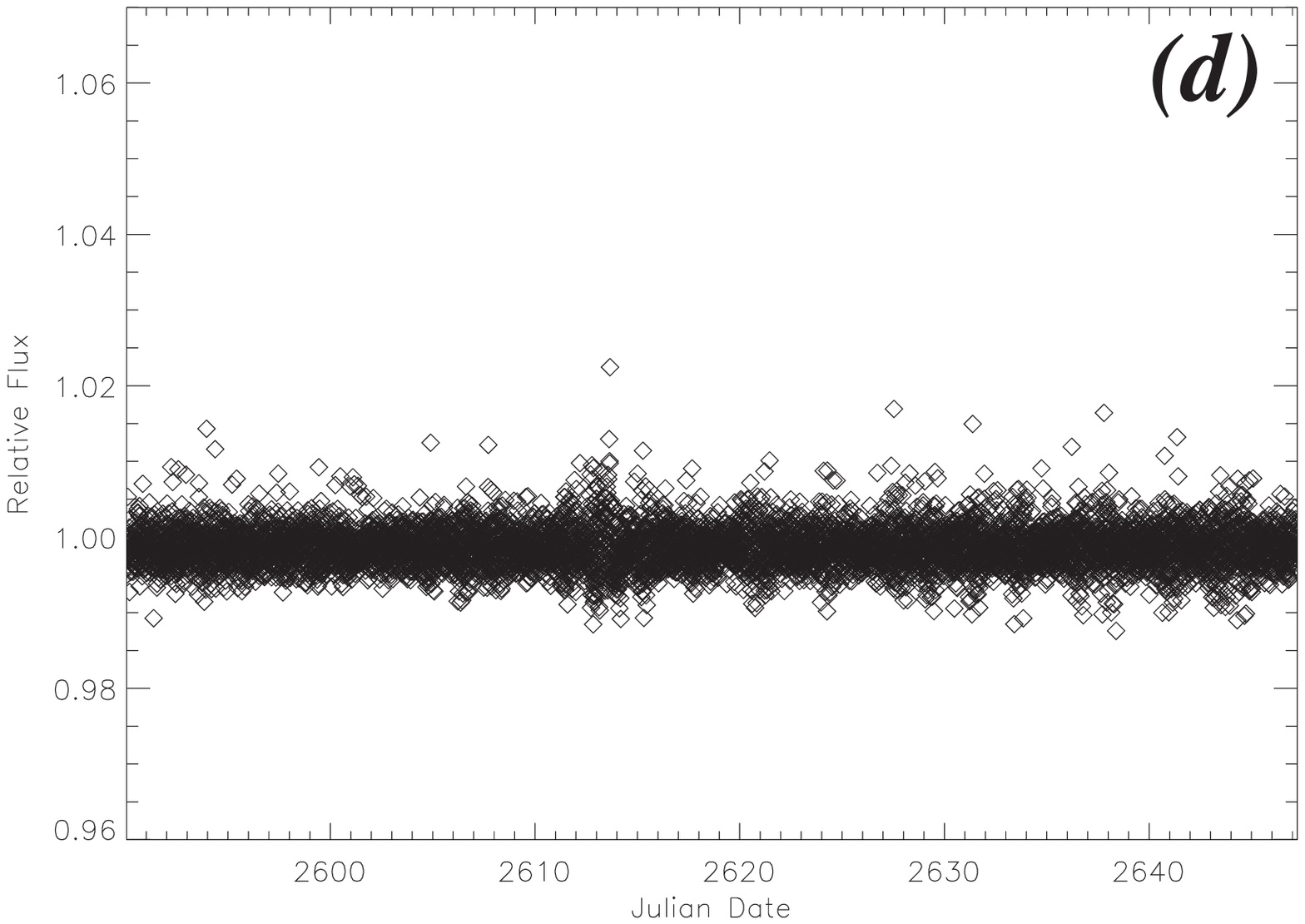} 

\caption{CoRoT012702789 red light curve and \textit{CDA} results. Raw data \textbf{(a)}, after 1 \textbf{(b)}, 3 \textbf{(c)}, and 5 \textbf{(d)} loops respectively. All jumps are removed.}
\label{fig3}
\end{figure*}
the procedure is completed, and $NFR_{final}$ is the final sub-light curve. The final step is to place all the 60 sub-light curves together. This produces the final light curve with which we are ready 
to search for exoplanets (Fig. \ref{fig3}). We employ of course many procedure loops, but if we use too many, \textit{CDA} begins to destroy the light curve because it is obvious that 
after a certain number of loops a ``saturation'' is reached in the procedure. To avoid this effect, we do not use the same loop number for each light curve. We calculate the standard deviation in each 
light curve after each loop and \textit{CDA} stops when the standard deviation begins to increases.

\subsection{Simulations}

To verify the functionality of CDA, we simulated CoRoT light curves as shown in Fig. \ref{sim1}. We simulated in particular a light curve in three filters (R,G,B), where jumps and trends appear at 
different times in each filter; a long-term trend was also included. In these light curves a transit pattern with period P=520 time units and a relative depth $\Delta Flux = 0.01$ was included. The 
transits were masked by the high noise. As can be seen in Fig. \ref{sim1}, all jumps were removed and the resulting output light curve shows some regions with higher noise and some others with lower 
noise, but this does not affect the real signal. By applying transit detection algorithms (e.g. box least squares - BLS \cite{2002A&A...391..369K}), the included transit pattern can also be detected.

\section{Results}

To illustrate the algorithm with real light curves, \textit{CDA} is applied to four CoRoT light curves, i.e., CoRoT0102702789, CoRoT0102874481, CoRoT0102741994, and CoRoT0102729260.

\subsection{The case of CoRoT0102702789}

In Fig. \ref{fig3}, we show the raw red light curve, which includes a trend and jumps, and the final light curve after applying \textit{CDA} with 5 loops. The light curve of CoRoT012702789 has one huge jump 
around $JD \sim 2614$ and many other smaller jumps. The $RSD_{R}$ value of the raw light curve is 5.048 and of the final light curve is 0.95. Table \ref{tab2} shows analytically the values of 
\textit{RSD} from the total light curves, in these 10 loops of each filter. The green filter has the minimum $RDS$ value, thus \textit{CDA} uses it as a base. The red filter on the other hand has 
the maximum value and we call it the target, but in principal \textit{CDA} defines different filters as either base or target in each loop. For this reason, in the first four loops the target is the red 
filter and base the green filter, then target changes to blue and green remains as base etc.; as already mentioned, the red light curve is the most common filter used to search for transits.

  \begin{table}

\begin{center}

      \caption[]{CoRoT01270289. This shows how \textit{RSD} varies in each loop. In the first four loops, red filter is the target and green the base. By loop five, this situation 
has changed: blue is the target now and green is the base. These values refers to the \textit{RSD} values of the full light curve of each filter.}

         \label{tab2}

        $\begin{array}{llll}

            \hline

            \noalign{\smallskip}

            Loop$ $No & RSD_{R} & RSD_{G} & RSD_{B} \\

            \noalign{\smallskip}

             \hline
            \noalign{\smallskip}

   \sharp1  &\textbf{ 5.0485} &    \textbf{0.9497}   &     1.0658     \\
   \sharp2  &\textbf{ 1.8632 }  &    \textbf{0.9497}   &     1.0658     \\
   \sharp3  & \textbf{1.0665}   &    \textbf{0.9497}   &     1.0658     \\
   \sharp4  & \textbf{0.9688 }  &    \textbf{0.9497}   &     1.0658     \\
   \sharp5  & 0.9688   &    \textbf{0.9497}   &     \textbf{0.9868 }    \\

            \hline\hline
         \end{array}$

\end{center}
   \end{table}

The example of CoRoT012702789 shows us how \textit{CDA} works and how it removes jumps from a distorted light curve.  As far as we can tell from out reconstructed light curve, there are no clear flares 
or transits in the light curve of  CoRoT012702789. The critical question at this point is how \textit{CDA} works if the raw light curve has real events such as transits.

\begin{figure}

\centering

\includegraphics[width=8cm,angle=0,clip=true]{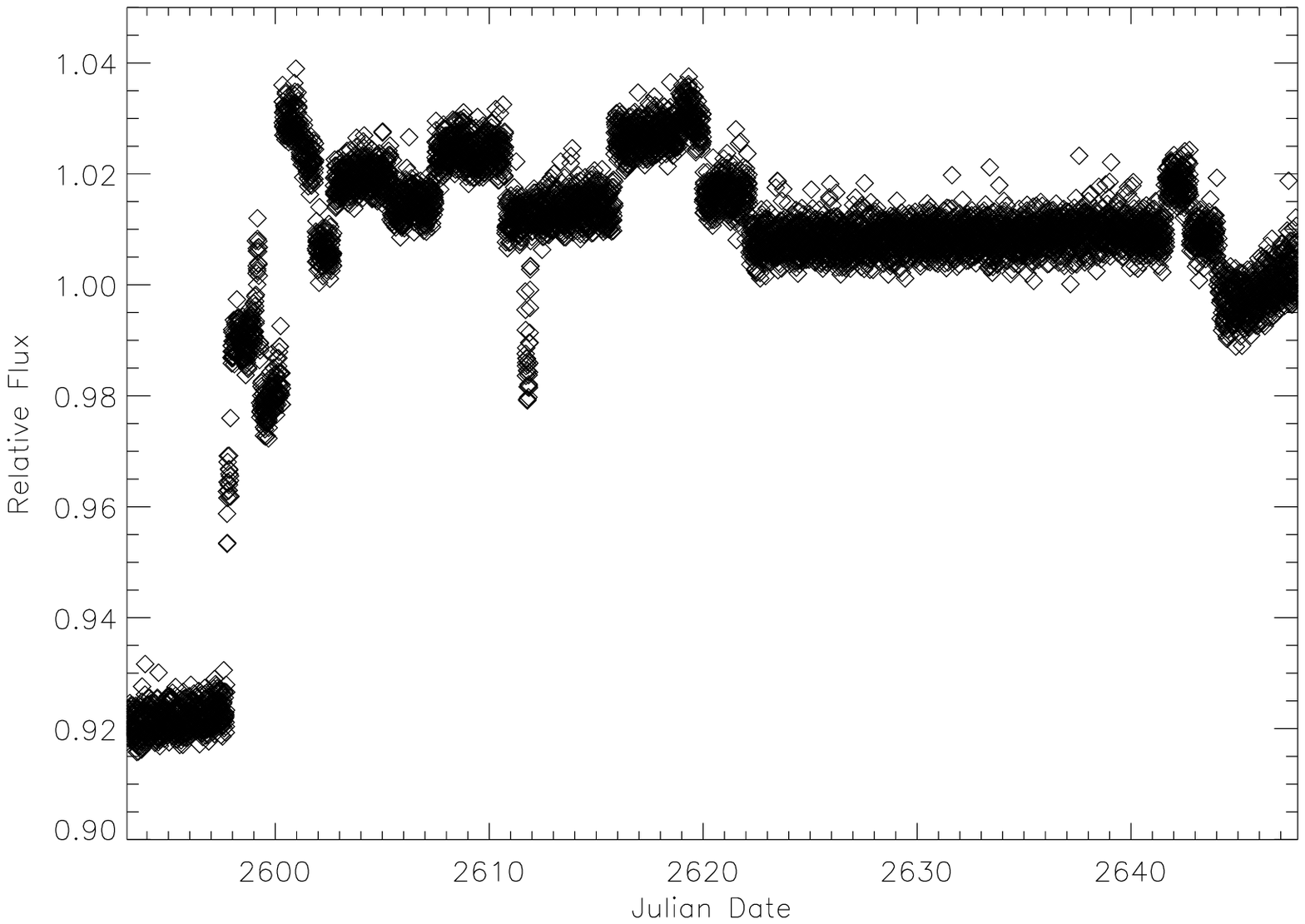}\\
\includegraphics[width=8cm,angle=0,clip=true]{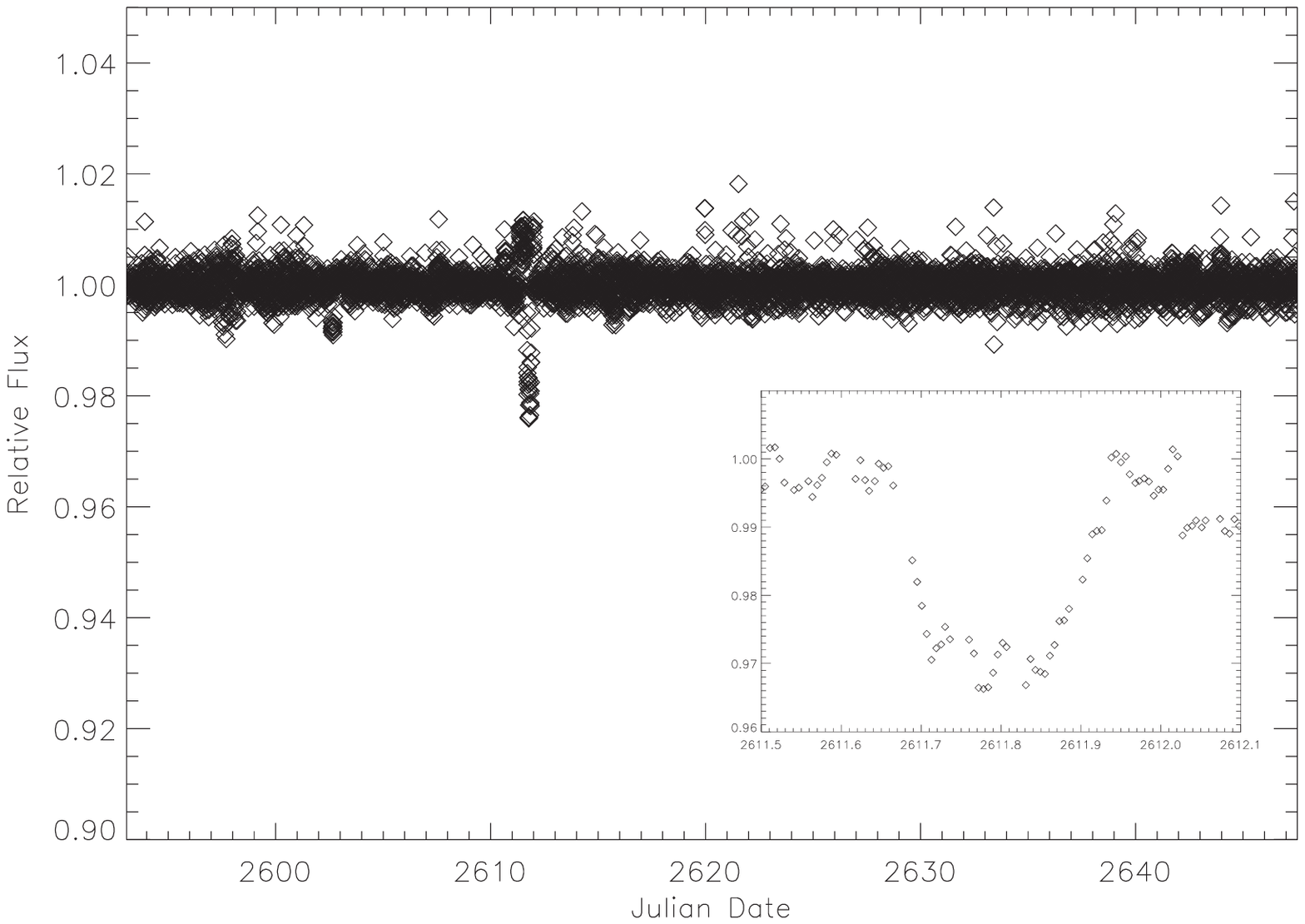}

\caption{CoRoT012874481 - red filter Top: Raw light curve. Bottom: The same light curve after \textit{CDA}. Jumps are removed and a clear transit appears. The subframe is a zoom-in plot.}
\label{fig4}

\end{figure}

\subsection{The case of CoRoT0102874481}

An even more extreme case is CoRoT0102874481, the light curve of which is affected by many jumps. The raw (red) light curve of CoRoT0102874481 is shown in Fig. \ref{fig4}. In the raw data, it is very 
difficult to distinguish real from instrumental events. As demonstrated in Fig. \ref{fig4}, \textit{CDA} corrects all the jumps except for a real transit around $JD \sim 2612$. The standard deviation 
before and after \textit{CDA} is 2203.13 and 336.44 ADUs, respectively. Only a small jump from green and blue filters remains at the end of light curve.\par
\begin{figure}

\centering

\includegraphics[width=8cm,angle=0,clip=true]{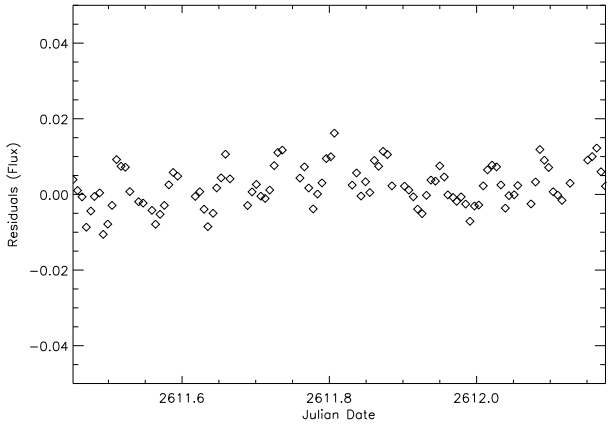}

\caption{CoRoT012874481 residuals before minus after \textit{CDA}. The signal from the real transit is not reduced by the algorithm.}
\label{fig5}
\end{figure}

Because this transit is the only transit in the light curve, we can determine neither the period nor the nature of the transiting object. Figure \ref{fig5} shows that \textit{CDA} does not reduce 
the depth of the transit, which is $\sim 0.036$. According to the CoRoT team\footnote{http://idoc-corot.ias.u-psud.fr}, the host star's spectral type is A0IV. Assuming
the typical radius and mass of this star to be $R_{s}=4.4R_{o}$ and $M_{s}=2.8M_{o}$ and assuming the transiting object to be a true exoplanet, we determine the planet's radius to be $R_{p}=4.28R_{J}$ 
by using the relation between radius and transit depth \citep{2003ApJ...585.1038S}

\begin{equation}
R_{p} = R_{s}\cdot \sqrt{\Delta Flux},
\label{eq8}
\end{equation}
where $R_{s}$ is the radius of the star and $R_{p}$ is the radius of the planet. From Kepler's $3^{rd}$ law, the semi-major axis of the orbit is $\alpha > 0.78AU$, because the period is $P>60$ days. 

\begin{figure}

\centering

\includegraphics[width=8cm,angle=0,clip=true]{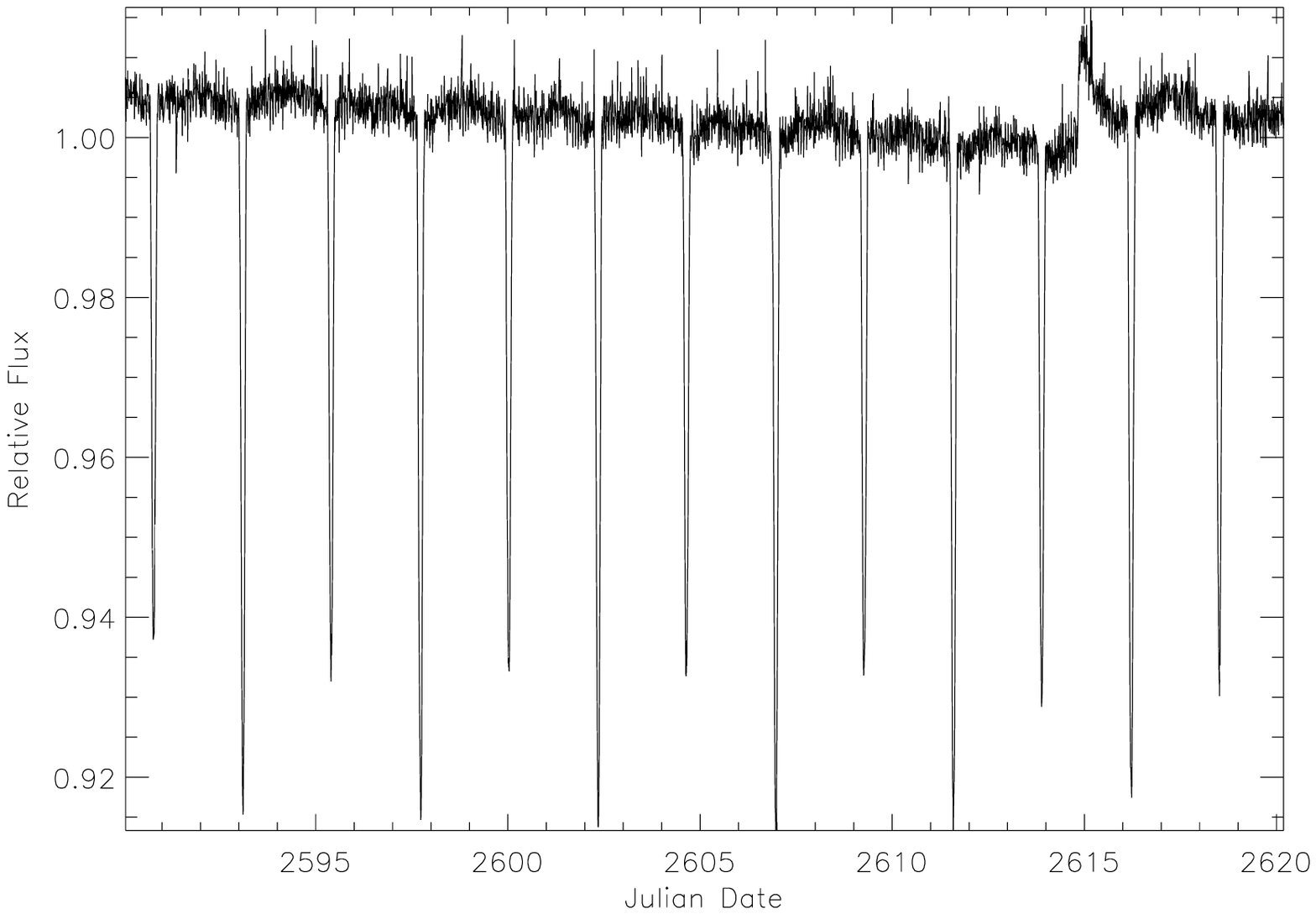}\\
\includegraphics[width=8cm,angle=0,clip=true]{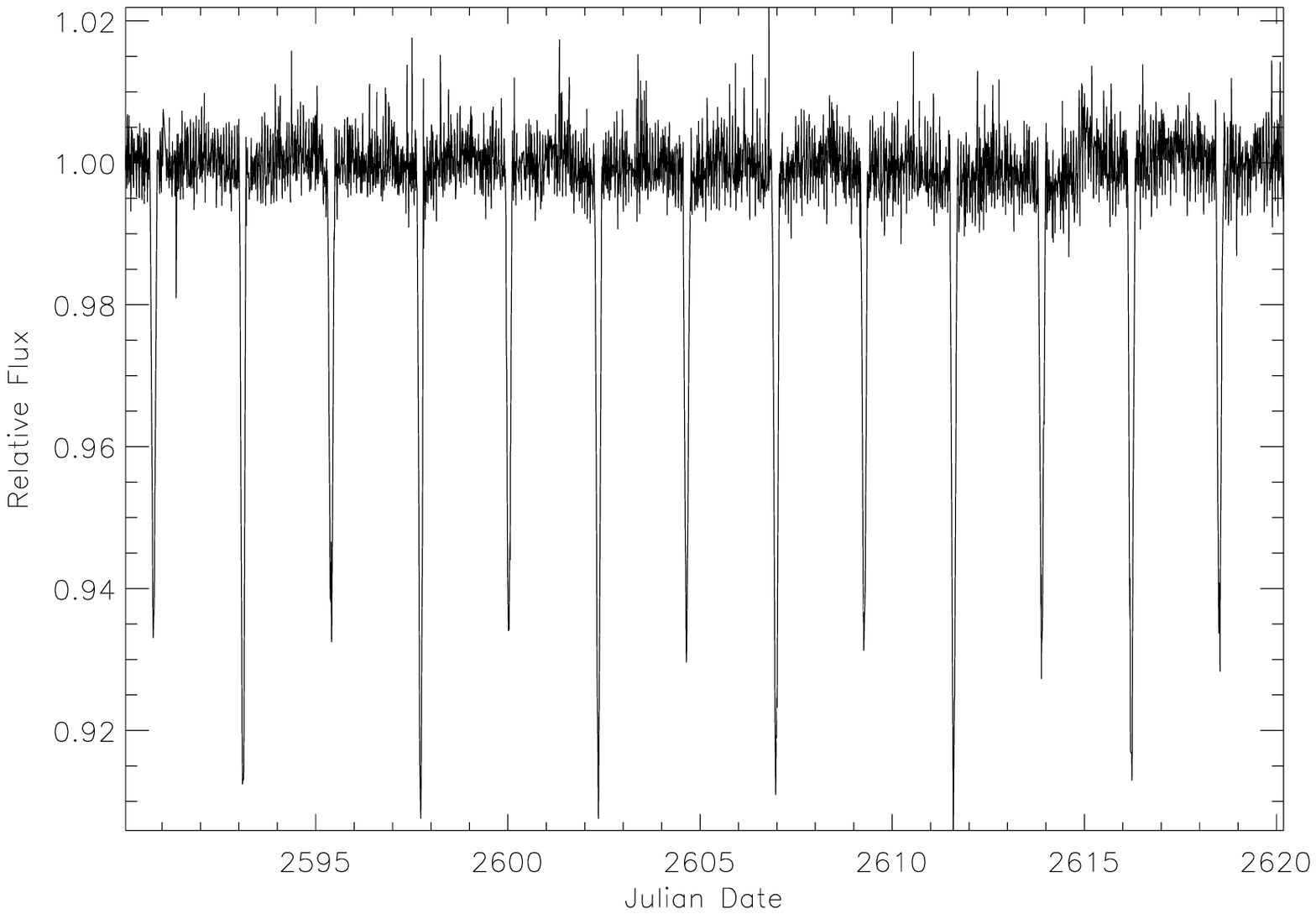}

\caption{CoRoT012741994 - red filter Up: Raw data. We have removed all the ``bad points''. The light curve contains one jump around $JD\sim 2615$ and a trend. Down: The same light 
curve after \textit{CDA}. The jump intensity is reduced. \textit{CDA} does not affect the transit depth.}

\label{fig6}
\end{figure}

\subsection{The case of CoRoT0102741994}

The source CoRoT0102741994 appears to be a binary system. In this example, our main interest is not to check whether \textit{CDA} can remove the jump but to check how the algorithm preserves the eclipses 
and the flux of the light curve. Figure \ref{fig6} shows how the algorithm converts the light curve. The light curve is affected by only a weak jump $(\Delta Flux \sim 1.25 \%)$ around $JD \sim 2615$. The 
flux depth of the primary and secondary eclipse is $9\%$ and $7\%$, respectively. In the top figure, is the light curve of the star before the applying \textit{CDA}. The two eclipses are obvious, while 
the bottom figure shows the light curve after application of \textit{CDA}. The jump is clearly removed completely. The depth of the primary and secondary eclipses are now $9.5\%$ and $6.5 \%$, 
respectively. As a general result, we can say that \textit{CDA} does not remove the real signal but corrects the jumps.

\subsection{The case of CoRoT0102729260}

\begin{figure}

\centering

\includegraphics[width=8cm,angle=0,clip=true]{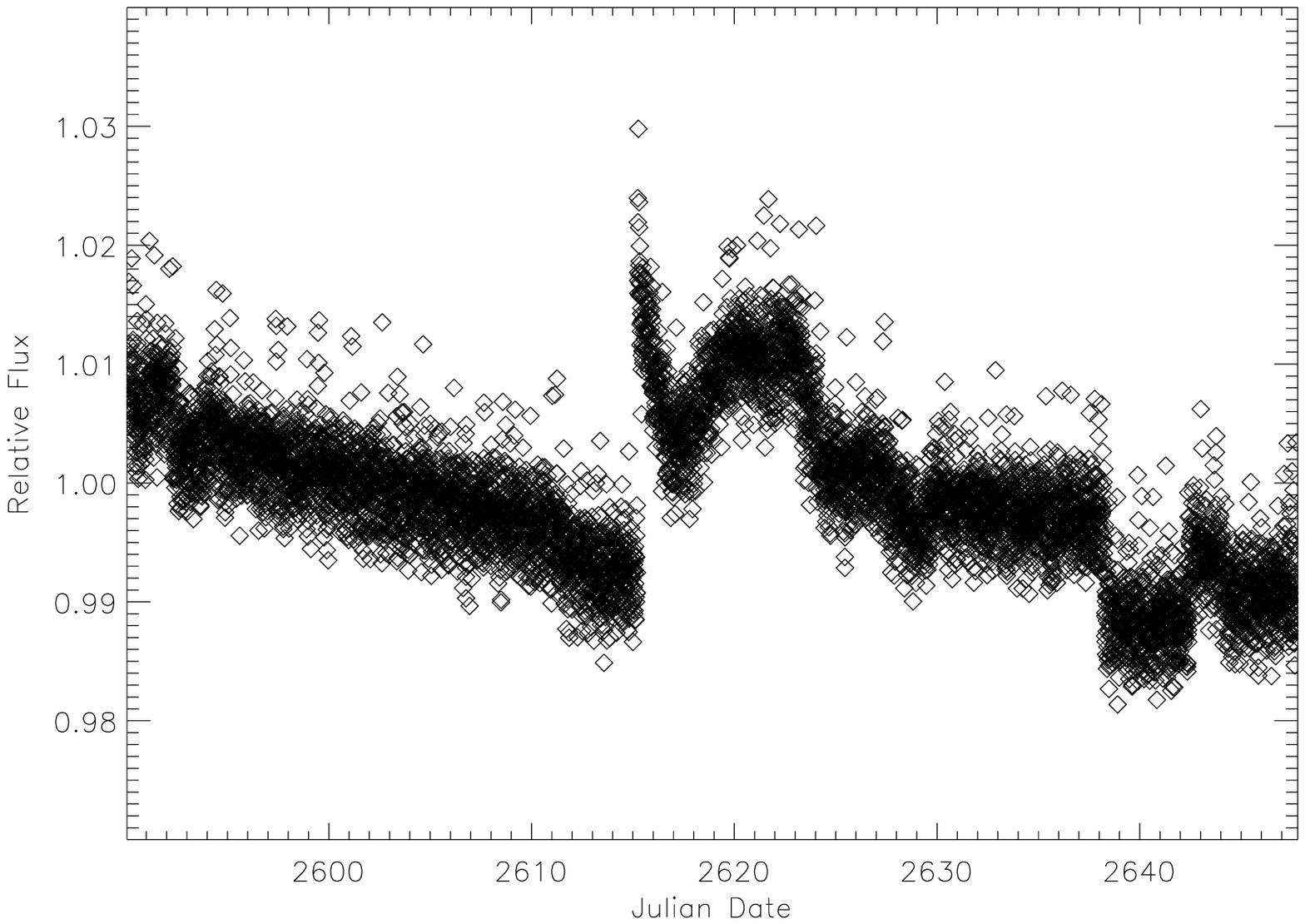}\\
\includegraphics[width=8cm,angle=0,clip=true]{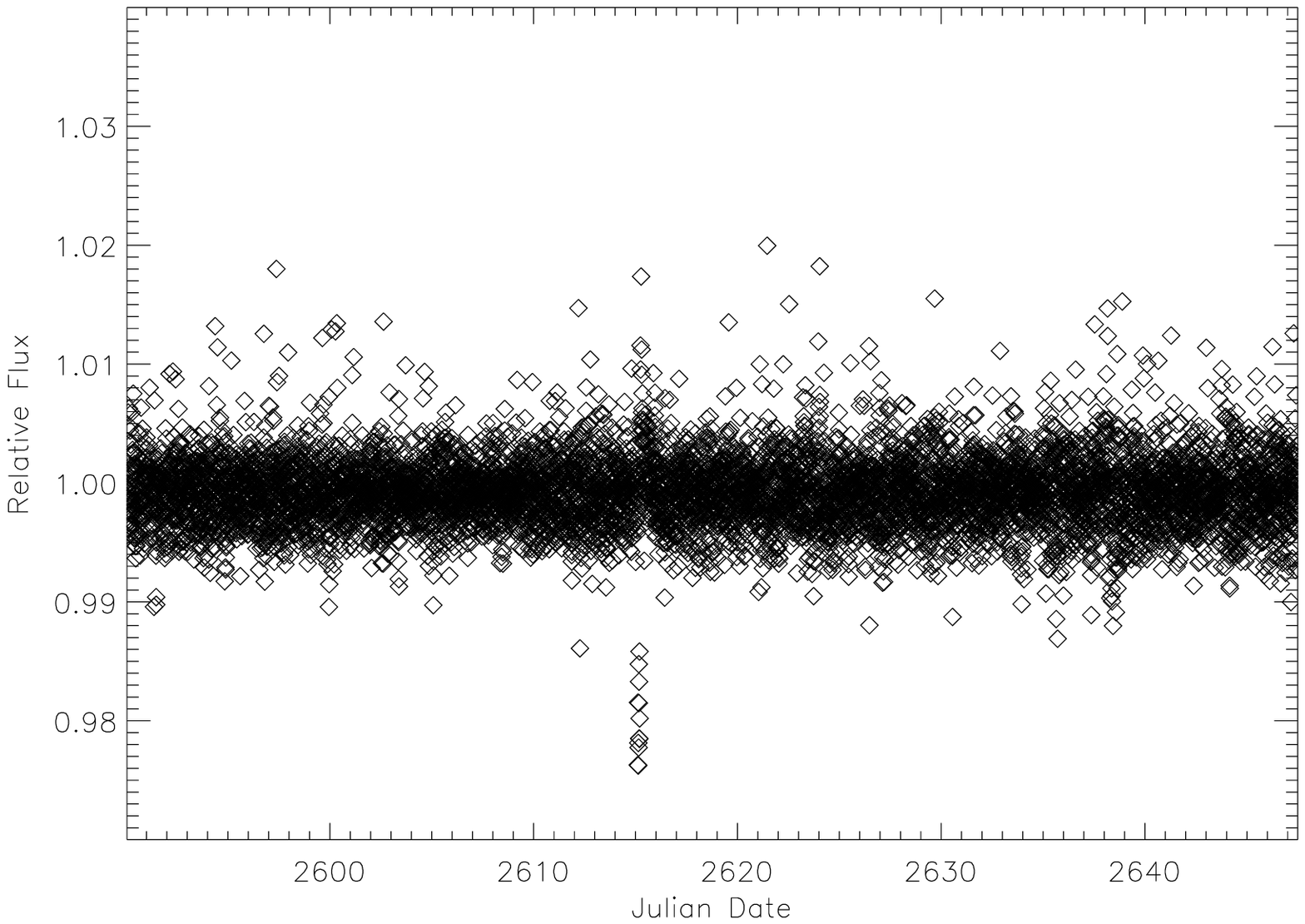}

\caption{CoRoT0102729260 - red filter. Up: Raw data before \textit{CDA}. Down: Final light curve after \textit{CDA}. The algorithm succeed to remove all the jumps and trends and improve the light curve 
enough to detect the ``concealed'' transit.}

\label{fig7}

\end{figure}

\begin{figure}

\centering

\includegraphics[width=8cm,angle=0,clip=true]{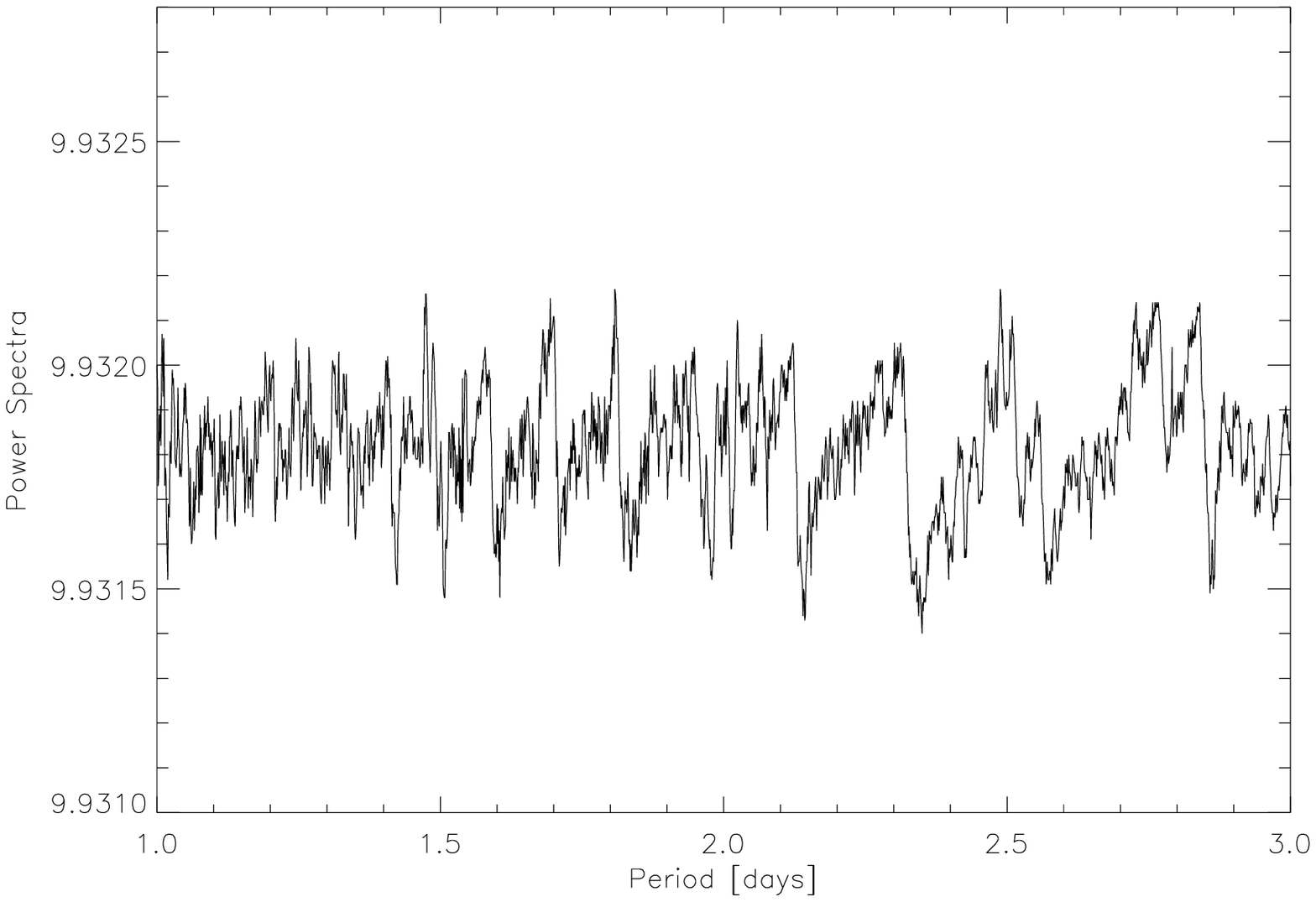}\\
\includegraphics[width=8cm,angle=0,clip=true]{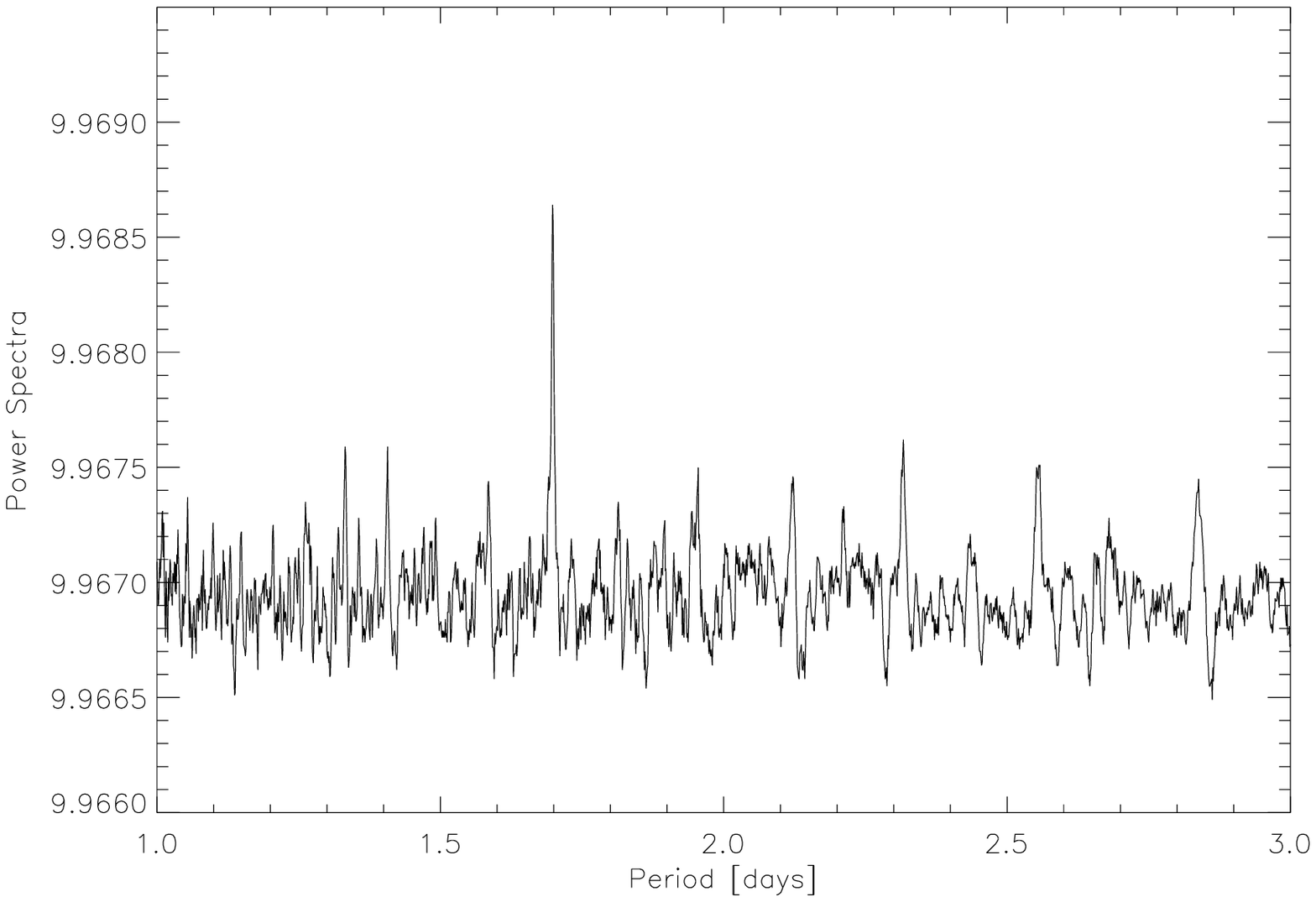}

\caption{CoRoT0102729260 - red filter. Up: Periodogramm of the raw light curve before \textit{CDA} without any obvious signal. Down: Same plot after \textit{CDA}. A clear periodic signal ($P \sim 1.698$) is detected.}

\label{spec}

\end{figure}

We find that the light curve of CoRoT0102729260 is a combination of trends and strong and weak jumps. The raw light curve of CoRoT0102729260 does not show any transits. We note that a transit 
detection algorithm such as BLS does not detect any transit event in this light curve (Fig. \ref{spec}, top panel). However, after applying \textit{CDA} to remove all jumps, we again implement BLS on 
the final light curve and a possible transit appears (Fig. \ref{fig7}, bottom panel). 

\begin{figure}

\centering

\includegraphics[width=8cm,angle=0,clip=true]{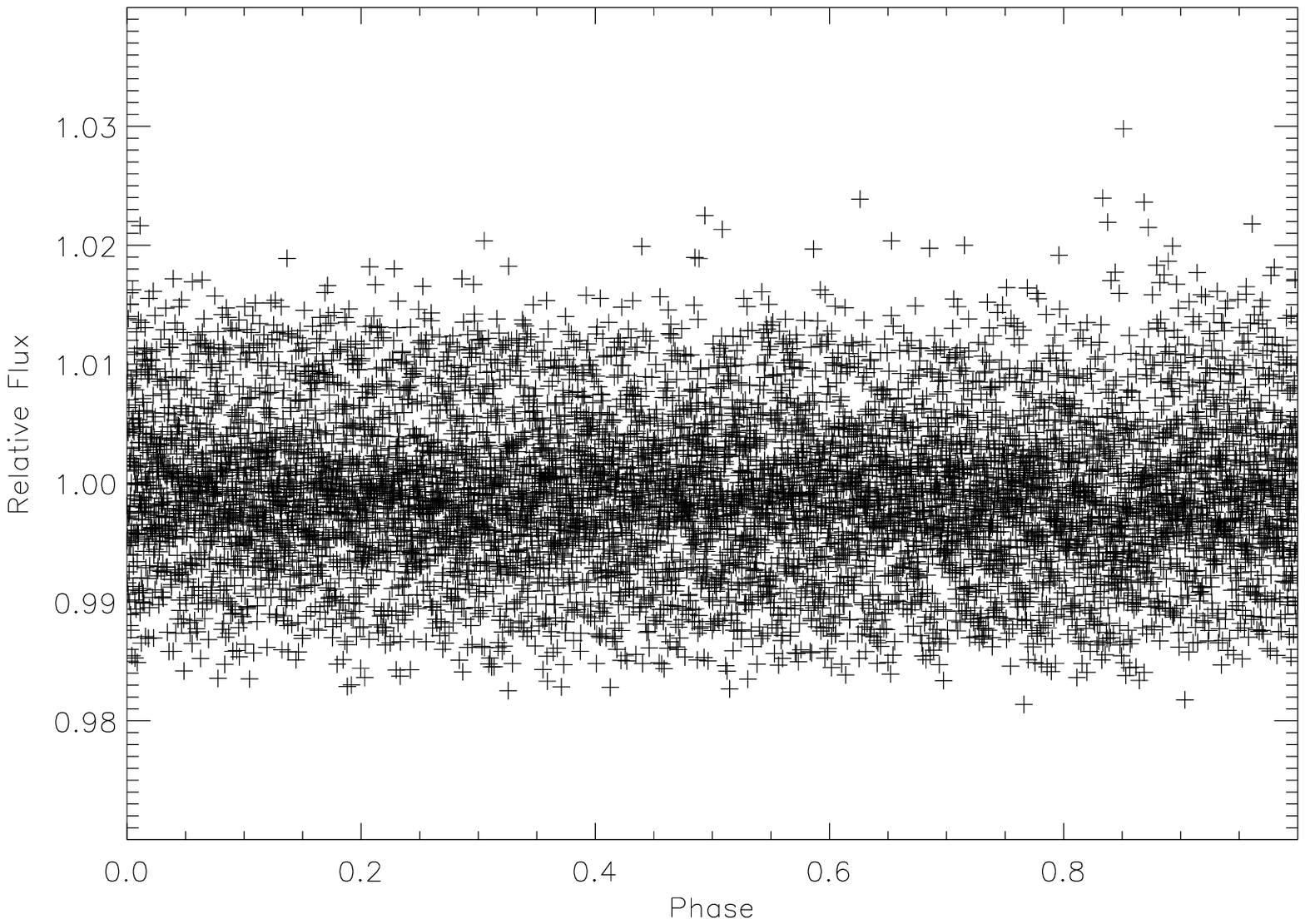}
\includegraphics[width=8cm,angle=0,clip=true]{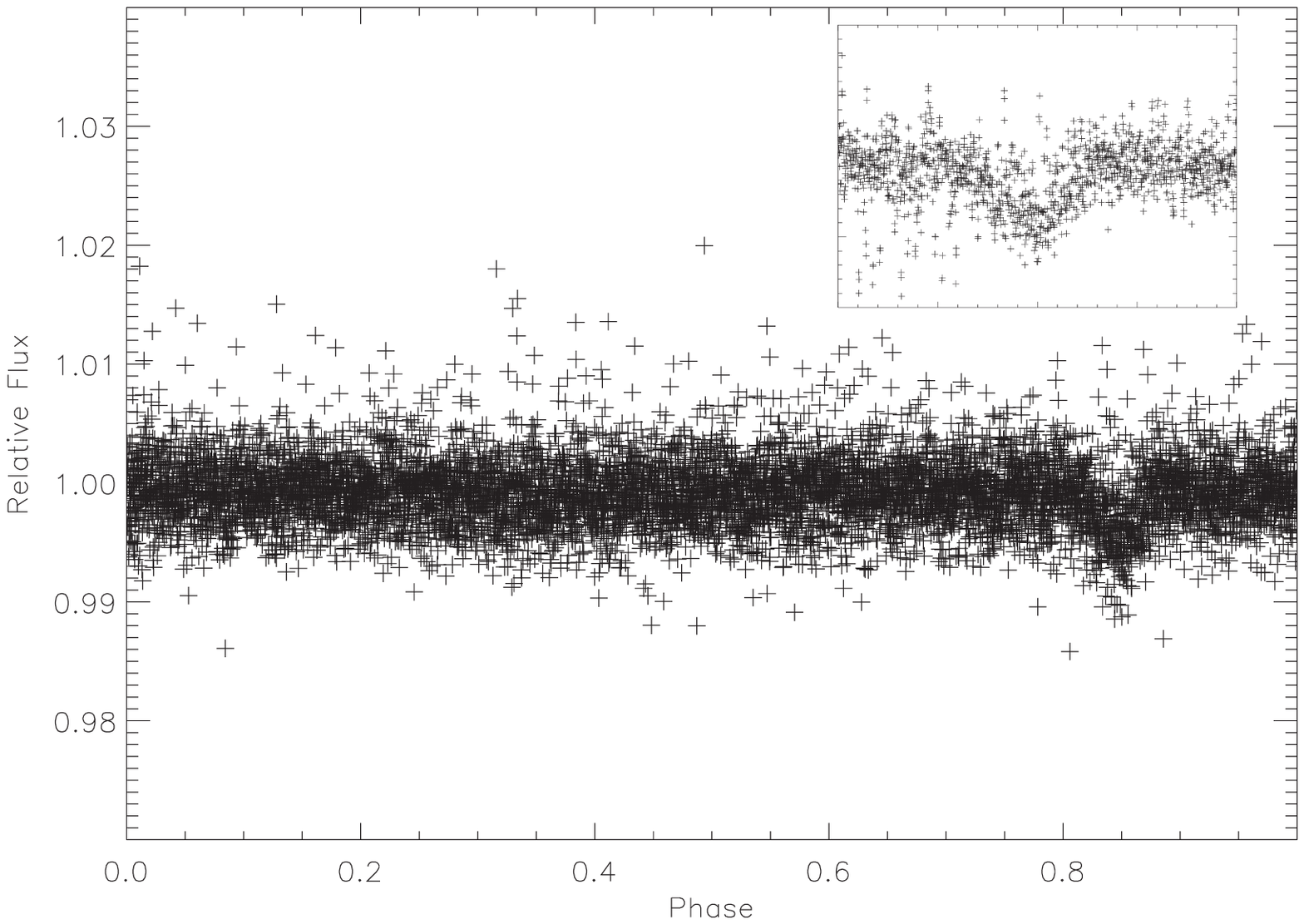}\\
\caption{CoRoT0102729260. Top: A phase-folded light curve before \textit{CDA}. Bottom: A phase-folded light curve after \textit{CDA}. }

\label{fig8}

\end{figure}

This transit is only detectable after applying \textit{CDA}, but not in the raw data. Our analysis of the phased light curve infers period of $P=1.6986$ days. The photometry by the CoRoT team provides 
some information about the parameter of the host star, which appears to be a main sequence star (G5V) of apparent brightness $m_{V} = 14.772$ mags. Assuming the spectral type to be correct, we can 
estimate the radius of the star to be $R_{s} \sim 0.91R_{o}$. With a transit depth of $\Delta Flux = 0.004$, we deduce a planetary radius of $R_{p}=6.27R_{E}$ applying Eq. \ref{eq8}. Figure \ref{fig8} 
shows the phase-folded light curves. Table \ref{tab3} also provides additional information about the system.

  \begin{table}

\begin{center}

      \caption[]{Physical parameters of CoRoT0102729260.}

         \label{tab3}

        $\begin{array}{ll}

            \noalign{\smallskip}

             \hline

            \noalign{\smallskip}

   Color$ $Index  & 0.752    \\
   Star$ $Radius$ $R_{s}          & 0.91R_{o}    \\
   Period & 1.6986$ $days      \\
   Planet$ $Radius$ $R_{p } & 6.27R_{E}     \\
   Depth$ $(Flux)  & 0.004    \\

            \hline\hline
         \end{array}$

\end{center}

   \end{table}

\section{Conclusions}

We have introduced and presented a method dubbed \textit{CDA} that removes instrumental artefacts from CoRoT data and demonstrated its usefulness in some practical applications. We emphasize that the 
\textit{CDA} algorithm can be used to prepares CoRoT data for any transit detection but should not be used for transit analysis because it can remove real signal. This is not of course a problem for the 
detection inasmuch as instrumental jumps affect far more the light curve. From our study of 1030 light curves in the first CoRoT field (IRao01), we found that only very few light curves have no 
instrumentally caused features and remain as they are, while the vast majority of light curves are appreciably improved. We have presented some examples that show how the algorithm affects the light 
curves. Our main conclusion is that instrumental jumps substantially affect the CoRoT light curves, making a transit detection in fainter stars impossible. \par
\begin{figure}

\centering

\includegraphics[width=8cm,angle=0,clip=true]{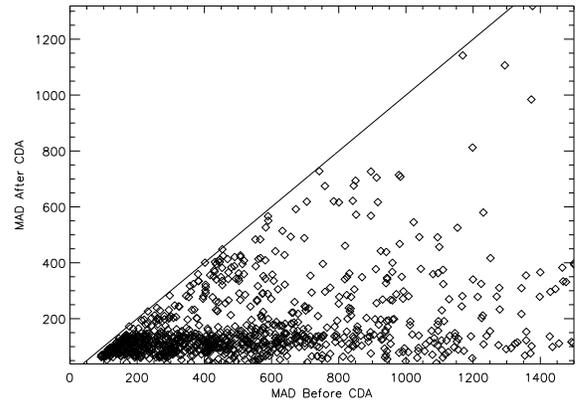}
\caption{Median absolute deviation (MAD) before and after \textit{CDA} using a 1030 light curve sample.}

\label{fig9}

\end{figure}

To illustrate how the algorithm affect the data of the full sample, we calculated the median absolute deviation (MAD) before and after applying \textit{CDA}. Figure \ref{fig9} shows the differences 
between the two procedures.\par
We prove our case with the example of CoRoT0102729260, a possible candidate exoplanet that is detected only after applying \textit{CDA} on the raw data.

\begin{acknowledgements}
DM was supported in the framework of the DFG-funded Research Training Group ''Extrasolar Planets and their Host Stars'' (DFG 1351/1).
\end{acknowledgements}

\bibliographystyle{aa}

\bibliography{aa}

\end{document}